\documentclass[prd,nofootinbib,amsfonts,notitlepage]{revtex4-1}
\usepackage[utf8]{inputenc}
\usepackage[T1]{fontenc}
\usepackage{hyperref}
\usepackage{graphicx,bm,amsmath,color}
\usepackage{bbold}
\usepackage{wasysym}
\usepackage{verbatim}
\usepackage{amssymb}
\usepackage{epstopdf}
\usepackage{animate}
\usepackage{xmpmulti}
\usepackage{centernot}
\usepackage{multirow}
\usepackage{tikz}
\usepackage{graphicx}
\usepackage{float}
\usepackage{mathtools}
\usepackage{comment}
\makeatletter

\newcommand{\be}{\begin{equation}}
\newcommand{\ee}{\end{equation}}
\newcommand{\beq}{\begin{eqnarray}}
\newcommand{\eeq}{\end{eqnarray}}
\newcommand{\ba}{\begin{align}}
\newcommand{\ea}{\end{align}}

\begin{document}

\title{Phenomenological consequences of a geometry in the cotangent bundle}
\author{J.J. Relancio}
\affiliation{Departamento de F\'{\i}sica Te\'orica and Centro de Astropartículas y Física de Altas Energías (CAPA),
Universidad de Zaragoza, Zaragoza 50009, Spain}
\author{S. Liberati}
\affiliation{SISSA, Via Bonomea 265, 34136 Trieste, Italy and INFN, Sezione di Trieste;\\ IFPU - Institute for Fundamental Physics of the Universe, Via Beirut 2, 34014 Trieste, Italy}
\email{relancio@unizar.es,liberati@sissa.it}

\begin{abstract}
A deformed relativistic kinematics can be understood within a geometrical framework through a maximally symmetric momentum space. However, when considering this kind of approach, usually one works in a flat spacetime and in a curved momentum space. In this paper, we will discuss a possible generalization to take into account both curvatures and some possible observable effects. We will first explain how to construct a metric in the cotangent bundle in order to have a curved spacetime with a nontrivial geometry in momentum space and the relationship with an action in phase space characterized by a deformed Casimir. Then, we will study within this proposal two different space-time geometries. In the Friedmann-Robertson-Walker universe, we will see the modifications in the geodesics (redshift, luminosity distance and geodesic expansion) due to a momentum dependence of the metric in the cotangent bundle. Also, we will see that when the spacetime considered is a Schwarzschild black hole, one still has a common horizon for particles with different energies, differently from a Lorentz invariance violation case. However, the surface gravity computed as the peeling off of null geodesics is energy dependent.   
\end{abstract}

\maketitle

\section{Introduction}
Due to the inconsistencies between general relativity (GR) and quantum field theory (QFT), a new theory that makes compatible both of them has been looked for several decades now. Examples of these attempts are string theory~\cite{Mukhi:2011zz,Aharony:1999ks,Dienes:1996du}, loop quantum gravity~\cite{Sahlmann:2010zf,Dupuis:2012yw}, supergravity~\cite{VanNieuwenhuizen:1981ae,Taylor:1983su}, or causal set \mbox{theory~\cite{Wallden:2013kka,Wallden:2010sh,Henson:2006kf}}. In most of these theories, a minimum length appears~\cite{Gross:1987ar,Amati:1988tn,Garay1995}, which is normally associated with the Planck length $\ell_P \sim 1.6\times 10^{-33}$\,cm. It is believed that this minimum length could mark somehow the transition to a ``quantum'' spacetime which replaces our concept of ``classical'' spacetime. 

If spacetime has a minimal length, there should be a modification of the special relativity (SR) symmetries, that characterized the classical spacetime, parametrized by a high energy scale (usually considered to be the Planck energy $\Lambda$)\footnote{See however~\cite{Rovelli:2002vp} for an alternative scenario.}. We can distinguish two different scenarios depending on how this modification is introduced. One can consider that some of the Poincaré group symmetries are broken, as in the Lorentz invariance violation (LIV) scenarios (see~\cite{Mattingly:2005re,Liberati2013} for a review), or one can have a deformation of these symmetries, in such a way that there is still a relativity principle. This is what is considered in the doubly special relativity (DSR) framework~\cite{AmelinoCamelia:2008qg}, where the kinematics of SR are deformed: there is a modified (nonlinear) conservation law for momenta, a deformed dispersion relation and, in order to have a relativity principle, a modification of the usual Lorentz invariance that makes compatible the two previous ingredients. This kind of deformation is usually carried out by a mathematical machinery called Hopf algebras~\cite{Majid:1995qg}, where the example of $\kappa$-Poincaré~\cite{Majid1994} is one of the most studied deformations of the Poincaré algebra.   
   
A duality between spacetime and momentum space was proposed by Born in the 30's~\cite{Born:1938}, considering the possibility that, if a curved spacetime describes GR, maybe a curved momentum space could represent a quantum gravity theory (when a curvature of spacetime is also present). This idea was formulated in order to avoid the ultraviolet divergences appearing in QFT, but recently has been considered again as a way to go beyond SR. In fact, it has been suggested in Refs.~\cite{AmelinoCamelia:2011bm,Lobo:2016blj} a relationship between a modified kinematics and a curved momentum space, which has been understood deeper in Ref.~\cite{Carmona:2019fwf}. From the algebraic point of view, in the particular case of the $\kappa$-Poincaré Hopf algebra~\cite{Lukierski:1991pn}, the associated $\kappa$-Minkowski noncommutative spacetime~\cite{Majid1994} allows us to deduce a de Sitter geometry for momentum space~\cite{KowalskiGlikman:2002jr}.

In order to study the possible consequences on spacetime of deviations from Lorentz invariance, there had been several papers studying Finsler geometries~\cite{Kostelecky:2011qz,Barcelo:2001cp,Weinfurtner:2006wt},  a generalization of a Riemannian geometry in which the metric can depend on the velocities (this is particular case of Lagrange space geometries~\cite{2012arXiv1203.4101M}). For example, in Ref.~\cite{Hasse:2019zqi} the redshift in Friedmann-Robertson-Walker and spherically symmetric metrics has been studied. Also, in Ref.~\cite{Stavrinos:2016xyg} the modified Raychaudhuri's equation has been developed for the Finslerian case. But in these works the modification of the metric is not related with a deformed kinematics in the DSR context. 

In the DSR framework, Finsler spacetimes have been studied for flat spacetime~\cite{Girelli:2006fw,Amelino-Camelia:2014rga}, and also for curved spacetimes~\cite{Letizia:2016lew}. In those papers it was shown that a deformed dispersion relation produce a velocity dependence on the metric. The main difference with the LIV scenario is that in this framework, nonlinear Lorentz transformations are implemented in order to make the metric invariant. A different approach was used in Ref.~\cite{Barcaroli:2015xda}, where the modification is carried out by Hamiltonian geometry (see also Ref.~\cite{2012arXiv1203.4101M}). In this case, the metric is momentum dependent (the Hamiltonian version of a Lagrange space). Both Finsler and Hamiltonian geometries are particular realizations of geometries in the tangent and cotangent bundle respectively. The starting point in all of them is a deformed dispersion relation. But in DSR context, and in particular in Hopf algebra framework, there is a basis called ``classical basis'' of $\kappa$-Poincaré~\cite{Borowiec2010} in which the dispersion relation is the usual of SR. As from a geometrical and algebraical point of view different basis are equivalent, one should obtain the same result starting from different dispersion relations. But this is not what it is found in those papers: in all of them, if one considers this particular basis of $\kappa$-Poincaré, one obtains the same results than in SR. 

Here we are going to consider a general case in the cotangent bundle because as we will see, this is required in order to study a modification of a spacetime due to a de Sitter momentum space. We will check that our metric is independent on the choice of the space-time variables one uses but, as in GR, the results depend on the momentum basis (choice of coordinates on the fiber). Our approach is completely different from the works appearing in the literature since our starting point is a metric in the cotangent bundle instead of a deformed dispersion relation, but as we will see, there is a relationship between both approaches.

The paper is organized as follows. In Sec.~\ref{sec:metric} we will see how to construct a metric in the cotangent bundle that takes into account the modified kinematics of $\kappa$-Poincaré, checking that there are isometries of the metric related translations and transformations that leave the momentum origin invariant for a fixed space-time point, which lead to the kinematics of  $\kappa$-Poincaré in absence of space-time curvature. This will manifest the presence of a nontrivial composition law, differencing the case of a deformed relativistic kinematics from the LIV scenario. Also, we will explain the main ingredients of the geometry in the cotangent bundle following~\cite{2012arXiv1203.4101M} that we will use in the paper, finding the modified Lie derivative in this context. Moreover, we will see the connection between the cotangent bundle metric formalism we will follow in the paper and the usual approach of considering an action in phase space with a deformed Casimir. In Sec.~\ref{sec:rw} and Sec.~\ref{sec:sch} we will see the phenomenological implications in the modified Friedmann-Robertson-Walker universe and in Schwarzschild black hole. In the first case we will study the modified geodesics, redshift, luminosity distance and the congruence of geodesics that takes into account the momentum dependence of the metric. For the Schwarzschild metric, we will study the null geodesics finding that particles with different energies will still have the same horizon, in contrast with the LIV case~\cite{Kifune:1999ex,Dubovsky:2006vk}. Also, we will compute the surface gravity from the peeling of null geodesics, finding that it depends on the energy. Finally in Sec.\ref{sec:conclusions}, we will see the conclusions.

\section{Metric in the cotangent bundle}
\label{sec:metric}

In this section we will first review the main results of~\cite{Carmona:2019fwf}. We will expose how a deformed relativistic kinematics can be understood through a maximally symmetric momentum space, characterized by a metric $g^{\mu\nu}_k(k)$. A deformed relativistic kinematics is composed of a deformed composition law for the momenta $\oplus$, a deformed dispersion relation $C(k)$ and, in order to have a relativity principle, modified Lorentz transformations ${\cal J}^{\alpha\beta}$. In particular, if the momentum space is de Sitter, one can find the special case of $\kappa$-Poincaré when one considers the isometries of the metric as the composition law (translations) and the Lorentz transformations (Lorentz isometries). 

After this incipit, we will show a possible way to generalize the previous work taking into account the curvature of spacetime. This will lead us to a metric in the cotangent bundle, depending on momentum and space-time coordinates. In such metric we will see that, as in the flat space-time case, one can define momentum transformations (for a fixed point $x$) that leaves the form of the metric invariant. Six of them leave the origin invariant (which are related to the Lorentz transformations) and the other four do not (which are related to translations, i.e. the composition law).

Also, we will explain how to deal with a metric in the cotangent bundle depending in both momentum and space-time coordinates, finding the deformed Killing equation for such metric.  

Finally, we will compare the velocity computed through the action with a deformed Casimir and the velocity obtained from a metric, checking that both procedures give the same result. 

\subsection{Curved momentum space, flat spacetime}
In~\cite{Carmona:2019fwf} a proposal to derive a (relativistic) deformed kinematics from a geometry in maximally symmetric momentum space is given,  defining a deformed composition and transformation laws from the isometries of the momentum metric associated to translations and Lorentz respectively, and the deformed dispersion relation as the (square of the) distance from the origin to a point in momentum space. In particular, it was shown in that paper that, when the momentum space is de Sitter with the choice of coordinates in which the metric is\footnote{We use the signature convention where $\eta=(+,-,-,-)$.} 
\begin{equation}
g_{00}(k)\,=\,1\,,\qquad g_{0i}(k)\,=\,0\,, \qquad  g_{ij}(k)\,=\,\eta_{ij}\, e^{-2k_0/\Lambda}\,,
\label{eq:dS_metric}
\end{equation}
where $\Lambda$ plays the role of the high energy scale, one can obtain the kinematics of $\kappa$-Poincar\'e in the bicrossproduct basis~\cite{Lukierski:1991pn}. Through the tetrad of momentum space defining the momentum metric
\begin{equation}
g^k_{\mu\nu}(k)\,=\, \varphi^\alpha_\mu(k)\eta_{\alpha\beta}\varphi^\beta_\nu(k) \,,
\end{equation}
it is easy to obtain the composition law, i.e the four translations, by the following equation
\begin{equation}
\varphi^\mu_\nu(p \oplus q) \,=\,  \frac{\partial (p \oplus q)_\nu}{\partial q_\rho} \, \varphi_\rho^{\,\mu}(q)\,.
\label{eq:tetrad_composition}
\end{equation}

Since these transformations leave the tetrad invariant (and then the metric), they are isometries of the momentum metric. With this prescription to obtain the composition law, it can be shown~\cite{Carmona:2019fwf} that the only compatible kinematics is $\kappa$-Poincaré (note that the composition law is associative by construction). 

With the choice of the tetrad leading to metric~\eqref{eq:dS_metric}
\begin{equation}
\varphi^0_0(k)\,=\,1\,,\qquad \varphi^0_i(k)\,=\,\varphi^i_0(k)\,=\,0\,,\qquad \varphi^i_j(k)\,=\,\delta^i_j e^{-k_0/\Lambda}\,,
\label{eq:dS_tetrad_p}
\end{equation}
the composition law is
\be
(p\oplus q)_0 \,=\, p_0 + q_0, \quad\quad\quad
(p\oplus q)_i \,=\, p_i + q_i e^{- p_0/\Lambda}\,.
\label{kappa-dcl}
\ee
 The modified Lorentz transformations are given  by the six isometries leaving invariant the origin:
\begin{equation}
\frac{\partial g^k_{\mu\nu}(k)}{\partial k_\rho} {\cal J}^{\alpha\beta}_\rho(k) \,=\,
\frac{\partial{\cal J}^{\alpha\beta}_\mu(k)}{\partial k_\rho} g^k_{\rho\nu}(k) +
\frac{\partial{\cal J}^{\alpha\beta}_\nu(k)}{\partial k_\rho} g^k_{\mu\rho}(k)\,,
\label{eq:cal(J)}
\end{equation}
where
\begin{equation}
 {\cal J}^{\alpha\beta}(k) \,=\,x^\mu {\cal J}^{\alpha\beta}_\mu(k)\,,
\end{equation}
is the Lorentz generator~\cite{Carmona:2019fwf}. From here one obtains 
 \be
{\cal J}^{0i}_0(k) \,=\, -k_i, \quad \quad \quad {\cal J}^{0i}_j(k)\,=\,  \delta^i_j \,\frac{\Lambda}{2} \left[e^{- 2 k_0/\Lambda} - 1 - \frac{\vec{k}^2}{\Lambda^2}\right] + \,\frac{k_i k_j}{\Lambda}\,.
\ee 
Once the latter is known, one can easily compute the Casimir defined as a function of momenta which is invariant under these transformations 
 \begin{equation}
\left\{C(k),{\cal J}^{\alpha\beta} \right\}\,=\, \frac{\partial C(k)}{\partial k_\mu}{\cal J}^{\alpha\beta}_\mu(k)\,=\,0\,,
\label{eq:casimir_J}
\end{equation}
getting
\be
C(k) \,=\, \Lambda^2 \left(e^{k_0/\Lambda} + e^{-k_0/\Lambda} - 2\right) - e^{k_0/\Lambda} \vec{k}^2  \,.
\ee
With all this we see that the ingredients of the deformed kinematics of $\kappa$-Poincar\'e in the bicrossproduct basis~\cite{Lukierski:1991pn} obtained through Hopf algebras can also be found from geometrical arguments~\cite{Carmona:2019fwf}.

\subsection{Curved momentum and space-time spaces}
In SR, one describes the motion of a free particle by the action  
\begin{equation}
S\,=\,\int{\dot{x}^\mu k_\mu-\mathcal{N} \left(C(k)-m^2\right)}\,,
\label{eq:SR_action}
\end{equation}
where $C(k)=k^\alpha \eta_{\alpha\beta }k^\beta$ is the SR dispersion relation and the dot represents the derivative with respect to $\tau$. One can obtain the geodesic motion in GR just rewriting Eq.\eqref{eq:SR_action} as
\begin{equation}
S\,=\,\int{\dot{x}^\mu k_\mu-\mathcal{N} \left(C(\bar{k})-m^2\right)}\,,
\label{eq:GR_action}
\end{equation}
where $\bar{k}_\alpha=\bar{e}^\nu_\alpha (x) k_\nu$, with $\bar{e}^\nu_\alpha(x)$ defined as the inverse of the tetrad of the space-time metric  $e^\nu_\alpha(x)$, satisfying
\begin{equation}
g^x_{\mu\nu}(x)\,= \, e^\alpha_\mu (x) \eta_{\alpha\beta} e^\beta_\nu (x)\,,
\label{eq:metric-st}
\end{equation}
while the dispersion relation is given by  
\begin{equation}
C(\bar{k})\,=\,\bar{k}^\alpha \eta_{\alpha\beta }\bar{k}^\beta\,=\,k^\mu g^x_{\mu\nu}(x) k^\nu\,.
\label{eq:cass_GR}
\end{equation}
One can check that the worldlines obtained through this action are the same that one would obtain in GR with the geodesics derived from the affine connection of the metric. 

In Ref.~\cite{AmelinoCamelia:2011bm} it was firstly proposed that the dispersion relation can be viewed as the squared distance from the origin to a point $k$ of the momentum space. In order to measure distances in momentum space, one can consider the line element  
\begin{equation}
d\sigma^2\,=\,dk_{ \alpha}g_k^{\alpha\beta}(k)dk_{ \beta}\,=\,dk_{ \alpha}\bar{\varphi}^\alpha_\gamma(k)\eta^{\gamma\delta}\bar{\varphi}^\beta_\delta(k)dk_{ \beta}\,,
\label{eq:line_m1}
\end{equation}
where $\bar{\varphi}^\alpha_\beta(p)$ is the inverse of $\varphi^\alpha_\beta(p)$ . Viewing the momentum space as a fiber of the space-time manifold, one compute such distance for a fixed space-time point (see chapter 4 of  Ref.~\cite{2012arXiv1203.4101M}). Then, if one considers that the transformation $k\rightarrow \bar{k}$ is the correct way to take into account a curvature in spacetime, the new momentum line element would be
\begin{equation}
d\sigma^2\,\coloneqq\,d \bar{k}_{\alpha}g_{\bar{k}}^{\alpha\beta}( \bar{k})d \bar{k}_{ \beta}\,=\,dk_{\mu}g^{\mu\nu}(x,k)dk_{\nu}\,,
\end{equation}
where in the second step we have used that the distance is carried along a fiber for a fixed space-time point. The tensor $g^{\mu\nu}(x,k)$ is constructed with the tetrad of spacetime and the original metric in momentum space. Explicitely,
\begin{equation}
g_{\mu\nu}(x,k)\,=\,\Phi^\alpha_\mu(x,k) \eta_{\alpha\beta}\Phi^\beta_\nu(x,k)\,,
\label{eq:cotangent_metric_tetrads}
\end{equation}
where 
\begin{equation}
\Phi^\alpha_\mu(x,k)\,=\,e^\lambda_\mu(x)\varphi^\alpha_\lambda(\bar{k})\,.
\label{eq:tetrad_cotangent}
\end{equation}
Now we can check that this metric is invariant under space-time diffeomorphisms, as in GR.
As this is a tetrad, a canonical transformation in phase space  $(x, k) \to (x',k')$ of the kind
\begin{equation}
x'^\mu\,=\,f^\mu(x)\,,\qquad k'_\mu\,=\,\frac{\partial x^\nu}{\partial x'^\mu} k_\nu\,,
\label{eq:canonical_transformation}
\end{equation} 
in such that for any nonlinear change of space-time variables, i.e for any set of functions $f_\mu$ of the space-time variables, the tetrad Eq.~\eqref{eq:tetrad_cotangent} will transform as
\begin{equation}
\Phi'^\mu_\rho(x',k')\,=\, \frac{\partial x^\nu}{\partial x'^\rho} \Phi^\mu_\nu(x,k)\,,
\label{eq:tetrad_canonical_transformation}
\end{equation}
because
\begin{equation}
\frac{\partial x^\mu}{\partial x'^\rho}  e^\lambda_\mu(x)\varphi^\alpha_\lambda(\bar{k})\,=\,e'^\kappa_\rho(x')\varphi'^\alpha_\kappa(\bar{k}')\,,
\label{eq:tetrad_transformation_fragmented}
\end{equation}
where we have used standard transformation law for the tetrad of spacetime
\begin{equation}
\bar{e}'^\nu_\mu(x')\,=\,\frac{\partial x'^\nu}{\partial x^\rho}\bar{e}^\rho_\mu(x)\,,
\end{equation} 
and then, the bared variables are independent of the choice of spatial coordinates 
\begin{equation}
\bar{k}'_ \mu\,=\,k'_\nu \bar{e}'^\nu_\mu(x')\,=\,\frac{\partial x^\sigma}{\partial x'^\nu}k_\sigma \frac{\partial x'^\nu}{\partial x^\rho} \bar{e}'^\rho_\mu(x')\,=\,k_\nu \bar{e}^\nu_\mu(x) \,=\,\bar{k}_ \mu\,.
\end{equation} 
Also, we consider that the momentum space tetrad do not change under such transformation.

In the following, we will prove that with the definition of the new momentum metric given in this work, starting with a momentum space metric, we can still define momentum transformations for a fixed space-time point $x$ that leave invariant the form of the metric, taking into account the curvature of the spacetime (the generalization to curved spacetime of the results obtained in~\cite{Carmona:2019fwf}): there are still 10 momentum isometries of the metric that correspond to the four translations  and six transformations that leave the origin invariant (the point in phase space $(x,0)$), and we can identify the squared distance from a point in the momentum space to the origin as the deformed dispersion relation. In Appendix~\ref{sec:appendix} it is shown that when the starting momentum space metric is of constant curvature (maximally symmetric space), the momentum scalar of curvature given by the contraction of Eq.~\eqref{eq:Riemann_p} is also constant.  Then, for a momentum metric with a dependence in space-time coordinates constructed with our procedure, we see that the fact that the original momentum space is maximally symmetric leads to a constant momentum scalar of curvature, and that we can also find 10 momentum isometries (momentum transformations for a fixed point in spacetime).

\subsubsection*{Modified translations}

As our starting point to take into account the curvature of the spacetime is to replace $k\rightarrow \bar{k}=\bar{e}k$, Eq.~\eqref{eq:tetrad_composition} should be generalized to 
\begin{equation}
\varphi^\mu_\nu(\bar{p} \oplus \bar{q}) \,=\,  \frac{\partial (\bar{p} \oplus \bar{q})_\nu}{\partial \bar{q}_\rho} \, \varphi_\rho^{\,\mu}( \bar{q})\,,
\label{eq:tetrad_composition2}
\end{equation}
where $p\rightarrow \bar{p}_\mu=\bar{e}_\mu^\nu(x) p_\nu$, $q\rightarrow \bar{q}_\mu=\bar{e}_\mu^\nu(x) q_\nu$. We can now define a modified composition ($\bar{\oplus}$) for a curved spacetime 
\be
(\bar{p} \oplus \bar{q})_\mu \,=\, \bar{e}_\mu^\nu(x) (p \bar{\oplus} q)_\nu\,.
\label{eq:composition_cotangent}
\ee
Then, one has 

\begin{equation}
\begin{split}
e^\tau_\nu(x)\varphi^\mu_\tau(\bar{p} \oplus \bar{q}) \,=&\,e^\tau_\nu(x)\frac{\partial (\bar{p} \oplus \bar{q})_\tau}{\partial \bar{q}_\sigma}\varphi_\sigma^{\,\mu}(\bar{q})\,=\,e^\tau_\nu(x) \bar{e}^\lambda_\tau(x) \,\frac{\partial (p \bar{\oplus} q)_\lambda}{\partial \bar{q}_\sigma}\varphi_\sigma^{\,\mu}(\bar{q})\\
=&\,\frac{\partial (p \bar{\oplus} q)_\nu}{\partial\bar{q}_\sigma} \varphi_\sigma^{\,\mu}(\bar{q})\,=\, 
\frac{\partial (p \bar{\oplus} q)_\nu}{\partial q_\rho}\frac{\partial q_\rho}{\partial\bar{q}_\sigma} \varphi_\sigma^{\,\mu}(\bar{q})\,=\,
\frac{\partial (p \bar{\oplus} q)_\nu}{\partial q_\rho} e_\rho^\sigma(x) 
\varphi_\sigma^{\,\mu}(\bar{q})\,,
\end{split}
\end{equation}
i.e.
\begin{equation}
\Phi^\mu_\nu(x,(p \bar{\oplus} q)) \,=\,  \frac{\partial (p \bar{\oplus} q)_\nu}{\partial q_\rho} \, \Phi_\rho^{\,\mu}(x,q)\,.
\label{eq:tetrad_cotangent_composition}
\end{equation}
This means that we can identify, for a fixed $x$, the isometries of this metric that leaves the form of the tetrad of the whole metric invariant as the deformed composition law when a curvature of spacetime is present, in the same way it was done in~\cite{Carmona:2019fwf}. 

By construction, the deformed composition law generators of  Eq.~\eqref{eq:tetrad_composition} form a group, so the composition law must be associative. By the same argument, the bared composition law must be associative. In fact, it is easy to see that if the composition law $\oplus$ is associative, the composition law $\bar{\oplus}$ is also associative. We define $\bar{r}=(\bar{k}\oplus \bar{q})$ and $\bar{l}=(\bar{p}\oplus \bar{k})$, and then we have  $r=(k \bar{\oplus} q)$ and $l=(p \bar{\oplus} k)$. Therefore,
\begin{equation}
(\bar{p}\oplus \bar{r})_\mu\,=\,\bar{e}^\alpha_\mu(p\bar{\oplus}r)_\alpha\,=\,\bar{e}^\alpha_\mu(p\bar{\oplus}(k\bar{\oplus}q))_\alpha\,,
\end{equation}
and 
\begin{equation}
(\bar{l}\oplus \bar{q})_\mu\,=\,\bar{e}^\alpha_\mu(l\bar{\oplus}q)_\alpha\,=\,\bar{e}^\alpha_\mu((p\bar{\oplus}k)\bar{\oplus}q)_\alpha\,,
\end{equation}
but as the $\oplus$ composition is associative, the following identity holds
\begin{equation}
(\bar{p}\oplus \bar{r})_\mu\,=\,(\bar{l}\oplus \bar{q})_\mu\,,
\end{equation}
and hence
\begin{equation}
(p\bar{\oplus}(k\bar{\oplus}q))_\alpha\,=\,((p\bar{\oplus}k)\bar{\oplus}q)_\alpha\,,
\end{equation}
so the $\bar{\oplus}$ is also associative. Thence, we have found that when one works in the cotangent bundle with a maximally symmetric momentum space, one can also define four momentum translations, which are also associative. 

\subsubsection*{Modified Lorentz transformations}

We can rewrite Eq.~\eqref{eq:cal(J)} replacing $k\rightarrow \bar{k}=\bar{e}k$
\be
\frac{\partial g^{\bar{k}}_{\mu\nu}(\bar{k})}{\partial \bar{k}_\rho} {\cal J}^{\beta\gamma}_\rho(\bar{k}) \,=\,
\frac{\partial{\cal J}^{\beta\gamma}_\mu(\bar{k})}{\partial \bar{k}_\rho} g^{\bar{k}}_{\rho\nu}(\bar{k}) +
\frac{\partial{\cal J}^{\beta\gamma}_\nu(\bar{k})}{\partial \bar{k}_\rho} g^{\bar{k}}_{\mu\rho}(\bar{k})\,.
\ee
From here, we have
\begin{equation}
\frac{\partial g^{\bar{k}}_{\mu\nu}(\bar{k})}{\partial k_\sigma}e^\rho_\sigma(x) {\cal J}^{\alpha\beta}_\rho(\bar{k}) \,=\,
\frac{\partial{\cal J}^{\alpha\beta}_\mu(\bar{k})}{\partial k_\sigma}e^\rho_\sigma(x) g^{\bar{k}}_{\rho\nu}(\bar{k}) +
\frac{\partial{\cal J}^{\alpha\beta}_\nu(\bar{k})}{\partial k_\sigma}e^\rho_\sigma(x) g^{\bar{k}}_{\mu\rho}(\bar{k})\,.
\end{equation}
Multiplying the previous equation by $e_\lambda^\mu(x)e_\tau^\nu(x)$ one obtains 
\begin{equation}
\frac{\partial g_{\lambda \tau}(x,k)}{\partial k_\rho} \bar{{\cal J}}^{\alpha\beta}_\rho(x,k) \,=\,
\frac{\partial\bar{{\cal J}}^{\alpha\beta}_\lambda(x,k)}{\partial k_\rho} g_{\rho\tau}(x,k) +
\frac{\partial\bar{{\cal J}}^{\alpha\beta}_\tau (x,k)}{\partial k_\rho}g_{\lambda\rho}(x,k)\,,
\end{equation}
where
\begin{equation}
 \bar{{\cal J}}^{\alpha\beta}_\mu(x,k) \,=\,e^\mu_\nu(x){\cal J}^{\alpha\beta}_\nu(\bar{k})\,.
\end{equation}
We see that $ \bar{{\cal J}}^{\alpha\beta}_\mu(x,k)$ are the new isometries of the metric leaving the momentum origin invariant for a fixed point $x$.  
\subsubsection*{Deformed dispersion relation}

Following our prescription, the generalization to Eq.~\eqref{eq:casimir_J} the previous equation when the spacetime is curved should be 
 \begin{equation}
\frac{\partial C(\bar{k})}{\partial \bar{k}_\mu}{\cal J}^{\alpha\beta}_\mu(\bar{k})\,=\,0\,.
\end{equation}
One can see the action of these transformations on the Casimir with the infinitesimal transformation parameters $\omega_{\alpha\beta}$:
\be
\delta C(\bar{k}) \,=\, \omega_{\alpha\beta} \frac{\partial C(\bar{k})}{\partial k_\lambda}\,\bar{{\cal J}}^{\alpha\beta}_\lambda(x,k) \,=\, \omega_{\alpha\beta} \frac{\partial C(\bar{k})}{\partial \bar{k}_\rho}\,\frac{\partial\bar{k}_\rho}{\partial k_\lambda}\,\bar{{\cal J}}^{\alpha\beta}_\lambda(x,k) 
=\,\omega_{\alpha\beta} \frac{\partial C(\bar{k})}{\partial \bar{k}_\rho}\,\bar{e}^\lambda_\rho(x)\,\bar{{\cal J}}^{\alpha\beta}_\lambda(x,k) \,=\, 
\omega_{\alpha\beta} \frac{\partial C(\bar{k})}{\partial \bar{k}_\rho}\,{\cal J}^{\alpha\beta}_\rho(\bar{k}) \,=\, 0\,,
\ee
where the last equality holds for a fixed point $x$. 

At the beginning of the section, in order to construct the momentum metric in presence of a curved spacetime, we have supposed that the replacement $k\rightarrow \bar{k}=\bar{e}k$ was a natural procedure to take into account the curvature of spacetime. With our prescription, we have found that if the Casimir $C(k)$ is the squared distance of the metric $g^k_{\mu\nu} (k)$ in momentum space from the origin to a point $k$,  $C(\bar{k})$ is the squared distance for a fixed point $x$ of the new momentum metric $g_{\mu\nu} (x,k)$ from the origin in momentum space to a point $k$. This means that our first assumption of considering that $C(\bar{k})$ is the deformed dispersion relation when the spacetime is curved, is consistent with how we constructed the momentum metric with a dependence on space-time coordinates, combining a curvature in spacetime with a curved momentum space.

\subsection{Main properties of geometry in the cotangent bundle}
We have seen that considering our approach we have obtained a metric in momentum space for a fixed point in spacetime. This metric can be considered as a metric in the cotangent bundle, taking into account a curvature in both momentum and space-time spaces, using the formalism given in Ch.4 of  Ref.~\cite{2012arXiv1203.4101M}. In this subsection, we are going to summarize the basic concepts and formulas we will use in the following. 

We define $H^\rho_{\mu\nu}$ as the affine connection of the metric in spacetime, in such a way that the covariant derivative of the metric vanishes 
\begin{equation}
g_{\mu\nu;\rho}(x,k)\,=\,  \frac{\delta g_{\mu\nu}(x,k)}{\delta x^\rho}-g_{\sigma\nu}(x,k)H^\sigma_{\rho\mu}(x,k)-g_{\sigma\mu}(x,k)H^\sigma_{\rho\nu}(x,k)\,=\,0\,,
\label{eq:cov_der}
\end{equation} 
where we use a new derivative 
\begin{equation}
\frac{\delta}{\delta x^\mu}\, \doteq \,\frac{\partial}{\partial x^\mu}+N_{\rho\mu}(x,k)\frac{ \partial}{\partial k_\rho}\,,
\label{eq:delta_derivative}
\end{equation}  
and  $N_{\mu\nu}(x,k)$ are the coefficients of the \textit{nonlinear connection} $N$ (also called horizontal distribution), supplementary to the vertical distribution $V$. The vertical distribution is generated by $\partial/\partial k_\mu$, while the horizontal one is constructed by $\delta/\delta x^\mu$. In GR, the coefficients of the \textit{nonlinear connection} are given by
\begin{equation}
N_{\mu\nu}(x,k)\, = \, k_\rho H^\rho_{\mu\nu}(x)\,.
\label{eq:nonlinear_connection}
\end{equation} 

Also, one can find the following relation between the metric and the affine connection
\begin{equation}
H^\rho_{\mu\nu}(x,k)\,=\,\frac{1}{2}g^{\rho\sigma}(x,k)\left(\frac{\delta g_{\sigma\nu}(x,k)}{\delta x^\mu} +\frac{\delta g_{\sigma\mu}(x,k)}{\delta  x^\nu} -\frac{\delta g_{\mu\nu}(x,k)}{\delta x^\sigma} \right)\,.
\label{eq:affine_connection_st}
\end{equation} 
The \textit{d-curvature tensor} is defined as~\cite{2012arXiv1203.4101M}
\begin{equation}
R_{\mu\nu\rho}(x,k)\,=\,\frac{\delta N_{\nu\mu}(x,k)}{\delta x^\rho}-\frac{\delta N_{\rho\mu}(x,k)}{\delta x^\nu}\,.
\label{eq:dtensor}
\end{equation} 
It represents the curvature of the phase space. It measures the integrability of spacetime, i.e. position space, as a subspace of the cotangent bundle and is defined as the commutator between the horizontal vector fields
\begin{equation}
\left\{ \frac{\delta}{\delta x^\mu}\,,\frac{\delta}{\delta x^\nu}\right\}\,=\, R_{\mu\nu\rho}(x,k)\frac{\partial}{\partial k_\rho}\,.
\end{equation} 
It can be seen that this tensor is 
\begin{equation}
R_{\mu\nu\rho}(x,k)\,=\,k_\sigma R^{*\sigma}_{\mu\nu\rho}(x,k)\,,
\end{equation} 
where 
\begin{equation}
R^{*\sigma}_{\mu\nu\rho}(x,k)\,=\,\left(\frac{\delta H_{\mu\nu}^\sigma(x,k)}{\delta x^\rho} -\frac{\delta H_{\mu \rho}^\sigma(x,k)}{\delta x^\nu} +H^\sigma_{\lambda\rho}(x,k)H_{\mu\nu}^\lambda(x,k)-H^\sigma_{\lambda\nu}(x,k)H_{\mu\rho}^\lambda(x,k)\right)\,.
\end{equation} 
In the GR case, $R_{\mu\nu\rho}(x,k)=k_\sigma R^{\sigma}_{\mu\nu\rho}(x)$, being $R^{\sigma}_{\mu\nu\rho}(x)$ the Riemann tensor. The horizontal bundle would be integrable if and only if $R_{\mu\nu\rho}=0$ (see Refs.~\cite{2012arXiv1203.4101M},\cite{Barcaroli:2015xda} for more details).

The affine connection in momentum space is 
\begin{equation}
C_\rho^{\mu\nu}(x,k)\,=\,\frac{1}{2}g_{\rho\sigma}\left(\frac{\partial g^{\sigma\nu}(x,k)}{\partial k_ \mu}+\frac{\partial g^{\sigma\mu}(x,k)}{\partial k_ \nu}-\frac{\partial g^{\mu \nu}(x,k)}{\partial k_ \sigma}\right)\,,
\label{eq:affine_connection_p}
\end{equation}
and then, we can also define the following covariant derivative 
\begin{equation}
v_{\nu}^{\,;\mu}\,=\,  \frac{\partial v_\nu}{\partial k_\mu}-v_\rho C^{\rho\mu}_\nu(x,k)\,.
\label{eq:k_covariant_derivative}
\end{equation}
The curvature tensor in position space is 
\begin{equation}
R^{\sigma}_{\mu\nu\rho}(x,k)\,=\,R^{*\sigma}_{\mu\nu\rho}(x,k)+C^{\sigma\lambda}_\mu (x,k)R_{\lambda\nu\rho}(x,k)\,,
\label{eq:Riemann_st}
\end{equation}
and the one in momentum space is 
\begin{equation}
S_{\sigma}^{\mu\nu\rho}(x,k)\,=\, \frac{\partial C^{\mu\nu}_\sigma(x,k)}{\partial k_\rho}-\frac{\partial C^{\mu\rho}_\sigma(x,k)}{\partial k_\nu}+C_\sigma^{\lambda\nu}(x,k)C^{\mu\rho}_\lambda(x,k)-C_\sigma^{\lambda\rho}(x,k)C^{\mu\nu}_\lambda(x,k)\,.
\label{eq:Riemann_p}
\end{equation}

One can define a line element in the cotangent bundle as  
\begin{equation}
\mathcal{G}\,=\, g_{\mu\nu}(x,k) dx^\mu dx^\nu+g^{\mu\nu}(x,k) \delta k_\mu \delta k_\nu\,, 
\end{equation}
where 
\begin{equation}
\delta k_\mu \,=\, d k_\mu - N_{\nu\mu}(x,k)\,dx^\nu\,. 
\end{equation}
In this way, a vertical path is characterized as a curve in the cotangent bundle with constant space-time coordinates and with the momentum satisfying the geodesic equation with the connection of the momentum space, i.e.
\begin{equation}
x^\mu\left(\tau\right)\,=\,x^\mu_0\,,\qquad \frac{d^2k_\mu}{d\tau^2}+C_\mu^{\nu\sigma}(x,k)\frac{dk_\nu}{d\tau}\frac{dk_\sigma}{d\tau}\,=\,0\,,
\end{equation} 
while an horizontal curve will be determined by
\begin{equation}
\frac{d^2x^\mu}{d\tau^2}+H^\mu_{\nu\sigma}(x,k)\frac{dx^\nu}{d\tau}\frac{dx^\sigma}{d\tau}\,=\,0\,,\qquad \frac{\delta k_\lambda}{\delta \tau}\,=\,\frac{dk_\lambda}{d\tau}-N_{\sigma\lambda} (x,k)\frac{dx^\sigma}{d\tau}\,=\,0\,.
\label{eq:horizontal_geodesics}
\end{equation} 
These are the same equations that hold in GR but, in this case, the affine connection $H^\mu_{\nu\sigma}(x,k)$ is a function depending not only on $x$ but on $k$.

\subsection{Modified Killing equation} 
In this subsection we will derive the modified Killing equation for a metric in the cotangent bundle. We can express the variation of the coordinates $x^\alpha$ along a vector field $\chi^\alpha$ as 
\begin{equation}
\left(x'\right)^\alpha\,=\,x^\alpha+\chi^\alpha \Delta\lambda\,,
\label{eq:x_variation}
\end{equation}
where $\lambda$ is the infinitesimal variation parameter. This variation of $x^\alpha$ reflects on $k_\alpha$ in the following way
\begin{equation}
\left(k'\right)_\alpha\,=\,k_\beta \frac{\partial x^\beta}{\partial x^{\prime \alpha}}\,=\,k_\alpha-\frac{\partial\chi^\beta}{\partial x^\alpha}k_\beta \Delta\lambda\,,
\end{equation}
since $k$ transforms as a covector. The general variation of a vector field $X^\alpha\left(x,k\right)$ will then be
\begin{equation}
\Delta X^\alpha\,=\,\frac{\partial X^\alpha}{\partial x^\beta} \Delta x^\beta+\frac{\partial X^\alpha}{\partial k_\beta} \Delta k_\beta\,=\,\frac{\partial X^\alpha}{\partial x^\beta}\chi^\beta \Delta\lambda-\frac{\partial X^\alpha}{\partial k_\beta}\frac{\partial\chi^\gamma}{\partial x^\beta}k_\gamma \,\Delta\lambda \,.
\label{eq:vector_variation}
\end{equation}
As in GR, in cotangent geometry we can obtain the Killing equation by imposing the line element invariance with respect to the variation along a vector field   $\chi^\alpha$
\begin{equation}
\Delta\left(ds^2\right)\,=\, \Delta (g_{\mu\nu}dx^\mu dx^\nu)\,=\, \Delta(g_{\mu\nu})dx^\mu dx^\nu+g_{\mu\nu} \Delta(dx^\mu) dx^\nu +g_{\mu\nu} \Delta(dx^\nu) dx^\mu\,=\,0\,.
\label{eq:line_variation}
\end{equation}
From Eq.\eqref{eq:vector_variation} we know that 
\begin{equation}
\Delta(g_{\mu\nu})\,=\,\frac{\partial g_ {\mu\nu}}{\partial x^\alpha} \chi^\alpha \Delta\lambda -\frac{\partial g_ {\mu\nu}}{\partial k_\alpha}\frac{\partial \chi^\gamma}{\partial x^\alpha}k_\gamma \,\Delta\lambda\,,
\end{equation}
while from  Eq.\eqref{eq:x_variation} we can obtain
\begin{equation}
\Delta(dx^\alpha)\,=\,d(\Delta x^\alpha)\,=\,d(\chi^\alpha \Delta\lambda)\,=\,\frac{\partial\chi^\alpha}{\partial x^\beta}dx^\beta\Delta\lambda\,.
\end{equation}
Therefore, Eq.\eqref{eq:line_variation} can be expressed as 
\begin{equation}
\Delta\left(ds^2\right)\,=\,\left(\frac{\partial g_ {\mu\nu}}{\partial x^\alpha} \chi^\alpha  -\frac{\partial g_ {\mu\nu}}{\partial k_\alpha}\frac{\partial \chi^\gamma}{\partial x^\alpha}k_\gamma \right) dx^\mu dx^\nu \Delta\lambda+ g_{\mu\nu}\left(\frac{\partial\chi^\mu}{\partial x^\beta}dx^\beta dx^\nu+\frac{\partial\chi^\nu}{\partial x^\beta}dx^\beta dx^\mu\right)\Delta\lambda\,,
\end{equation}
giving finally
\begin{equation}
\frac{\partial g_ {\mu\nu}}{\partial x^\alpha} \chi^\alpha  -\frac{\partial g_ {\mu\nu}}{\partial k_\alpha}\frac{\partial \chi^\gamma}{\partial x^\alpha}k_\gamma + g_{\alpha\nu}\frac{\partial\chi^\alpha}{\partial x^\mu}+ g_{\alpha\mu}\frac{\partial\chi^\alpha}{\partial x^\nu}\,=\,0\,,
\label{eq:killing}
\end{equation}
which is the same equation obtained in~\cite{Barcaroli:2015xda}. This can be rewritten in a covariant way taking into account the fact that $\chi^\alpha$ does not depend on $k$, and then 
\begin{equation}
\frac{\partial \chi^\alpha}{\partial x^\beta}\,=\, \frac{\delta\chi^\alpha}{\delta x^\beta}\,,
\end{equation}
so the previous equation becomes
\begin{equation}
0\,=\, \left(\frac{\delta g_{\mu\nu}}{\delta x^\alpha}-\frac{\partial g_{\mu\nu}}{\partial k_\rho}H^\gamma_{\rho\alpha}k_\gamma\right) g^{\alpha\beta}\chi_\beta-\frac{\partial g_ {\mu\nu}}{\partial k_\alpha}\frac{\delta\chi^\gamma}{\delta x^\alpha}k_\gamma+
 g_{\lambda\nu}\left(\frac{\delta g^{\lambda\alpha}}{\delta x^\mu}\chi_\alpha+ g^{\lambda\alpha} \frac{\delta \chi_\alpha}{\delta x^\mu} \right)+g_{\lambda\mu}\left(\frac{\delta g^{\lambda\alpha}}{\delta x^\nu}\chi_\alpha+ g^{\lambda\alpha} \frac{\delta \chi_\alpha}{\delta x^\nu} \right)\,,
\end{equation}
and using the definitions of the affine connection of Eq.~\eqref{eq:affine_connection_st} and covariant derivative of Eq.~\eqref{eq:cov_der} one finds
\begin{equation}
\mathcal{L}_\chi g_{\mu\nu}\,=\,\chi_{\nu;\mu}+\chi_{\mu;\nu}-\frac{\partial g_ {\mu\nu}}{\partial k_\alpha} \chi^\gamma_{\,;\alpha} k_\gamma\,=\,0\,.
\label{eq:lie_metric}
\end{equation}
Also we can find the modified Lie derivative for a contravariant vector
\begin{equation}
\mathcal{L}_\chi u^\mu\,=\,\chi^\nu u^\mu_{\,;\nu}-u^\nu \chi^\mu_{\,;\nu}-\frac{\partial u^\mu}{\partial k_\alpha} \chi^\gamma_{\,;\alpha} k_\gamma\,.
\label{eq:lie_vec}
\end{equation}

\subsection{Relationship between metric and action formalisms}

Let us consider the line element in momentum space. One can find a simple and useful relation between the distance and the metric for a Riemannian manifold~\cite{Bhattacharya2012RelationshipBG} 
 \begin{equation}
\frac{\partial D(0,k)}{\partial k_\mu}\,=\,\frac{k_\nu g^{\mu\nu}(k)}{\sqrt{k_\rho g^{\rho\sigma}(k) k_\sigma}}
\end{equation}
where $D(0,k)$ is the distance from a fixed point $0$ to $k$. This implies 
 \begin{equation}
\frac{\partial D(0,k)}{\partial k_ \mu}g_{\mu\nu}(k) \frac{\partial D(0,k)}{\partial k_ \nu}\,=\,1\,.
\end{equation}
In Ch.3 of~\cite{Petersen2006} it has been showed that this property also holds for the Minkowski space (inside the light cone and extended on the light cone by continuity) and hence, it is valid for any pseudo Riemannian manifold of dimension $n$ due to Whitney embedding theorem~\cite{Burns1985}, since they can be embedded in a Minkowski space of at most dimension $2n+1$. From this property, it is easy to obtain a simple relationship between the metric and the Casimir defined as the distance squared
 \begin{equation}
\frac{\partial C(k)}{\partial k_ \mu}g_{\mu\nu}(k) \frac{\partial C(k))}{\partial k_ \nu}\,=\,4 C(k)\,.
\label{eq:casimir_definition}
\end{equation}

From the action 
\begin{equation}
S\,=\,\int{\left(\dot{x}^\mu k_\mu-\mathcal{N} \left(C(k)-m^2\right)\right)d\tau}\,,
\label{eq:DSR_action}
\end{equation}
with a generic deformed Casimir, we can read that 
\begin{equation}
\dot{x}^\mu\,=\,\mathcal{N}\frac{\partial C(k)}{\partial k_\mu}\,,
\label{eq:velocity_action}
\end{equation}
being $\mathcal{N}=1/2m$ or $1$ when the curve is timelike or null respectively. 

Following the prescription of the previous subsection, we can consider the line element in spacetime to be 
\begin{equation}
ds^2\,=\, g_{\mu\nu}(k) dx^\mu dx^\nu\,.
\end{equation}
For the timelike case, we can chose the parameter of the curve to be $s$ and then 
\begin{equation}
1\,=\, \dot{x}^\mu g_{\mu\nu}(k) \dot{x}^\nu\,.
\end{equation}
Substituting Eq.~\eqref{eq:velocity_action} in the previous equation we find 
\begin{equation}
\left. \frac{1}{4 m^2} \frac{\partial C(k)}{\partial k_\mu} g_{\mu\nu}(k) \frac{\partial C(k)}{\partial k_\nu}\right\rvert_{C(k)=m^2}\,=\, \frac{1}{4 m^2} 4 m^2\,=\,1 \,,
\end{equation}
where we have used Eq.~\eqref{eq:casimir_definition}. If we consider a null geodesic, then 
\begin{equation}
0\,=\, \dot{x}^\mu g_{\mu\nu}(k) \dot{x}^\nu\,,
\end{equation}
and therefore, using Eq.~\eqref{eq:velocity_action} we find
\begin{equation}
\left. \frac{\partial C(k)}{\partial k_\mu} g_{\mu\nu}(k) \frac{\partial C(k)}{\partial k_\nu}\right\rvert_{C(k)=0}\,=\,0 \,,
\end{equation}
where again Eq.~\eqref{eq:casimir_definition} was used in the last step. We see that considering an action with a deformed dispersion relation and a momentum geometry where we identify the squared distance with the Casimir, leads us to the same results~\footnote{If instead of considering the squared distance one considers a function of it, one arrives to the same results just redefining the mass for the timelike curves. The null cases would be exactly the same.}. 

This is  also valid for the generalization we propose in this work considering a curved spacetime and momentum spaces. In this case, the relation of Eq.~\eqref{eq:casimir_definition} is generalized to
 \begin{equation}
\frac{\partial C(\bar{k})}{\partial \bar{k}_ \mu}g^{\bar{k}}_{\mu\nu}(\bar{k}) \frac{\partial C(\bar{k}))}{\partial \bar{k}_ \nu}\,=\,4 C(\bar{k})\,=\,\frac{\partial C(\bar{k})}{\partial k_ \mu}g_{\mu\nu}(x,k) \frac{\partial C(\bar{k}))}{\partial k_\nu}\,.
\label{eq:casimir_definition_cst}
\end{equation}
From the action 
\begin{equation}
S\,=\,\int{\dot{x}^\mu k_\mu-\mathcal{N} \left(C(\bar{k})-m^2\right)}
\label{eq:DGR_action}
\end{equation}
with the same deformed Casimir but depending on the bared momenta, we can read 
\begin{equation}
\dot{x}^\mu\,=\,\mathcal{N}\frac{\partial C(\bar{k})}{\partial k_\mu}\,,
\label{eq:velocity_action_curved}
\end{equation}
where again $\mathcal{N}=1/2m$ or  $1$ when the curve is timelike or null respectively. Then, we can trivially see that, with the generalization considered here, we observe the same relationship between the action and metric formalisms.

\section{Friedmann-Robertson-Walker metric}
\label{sec:rw}

Now we can study different models for spacetime with a de Sitter momentum space. In this section, we will start by computing the momentum dependence of velocity in the case of photons in two different ways for the Friedmann-Robertson-Walker metric. We will see that the results are the same obtained from the variation of the action Eq.~\eqref{eq:GR_action} and computed through the line element of the metric, which is in agreement with what we have found in the previous section. Moreover, we will obtain the evolution of momenta as a function of time. We will also study some phenomenological aspects related with the Friedmann-Robertson-Walker universe. 

In order to construct the metric in the cotangent bundle, we choose the tetrad of de Sitter momentum space of Eq.~\eqref{eq:dS_tetrad_p}, while for the space-time metric, we choose the tetrad to be
\begin{equation}
e^0_0(x)\,=\,1\,,\qquad e^0_i(x)\,=\,e^i_0(x)\,=\,0\,,\qquad e^i_j(x)\,=\,\delta^i_j R(x_0)\,,
\label{eq:RW_tetrad_st}
\end{equation}
where $R(x_0)$ is the scale factor. With these tetrads we are now able to construct the metric of the cotangent bundle from Eq.~\eqref{eq:cotangent_metric_tetrads}, obtaining 
\begin{equation}
g_{00}(x,k)\,=\,1\,,\qquad g_{0i}(x,k)\,=\,0\,, \qquad  g_{ij}(x,k)\,=\,\eta_{ij}\, R^2(x^0) e^{-2k_0/\Lambda}\,.
\label{eq:RW_metric}
\end{equation}

For this metric, one can see from Eq.~\eqref{eq:Riemann_p} that the scalar of curvature in momentum space is constant $S=12/\Lambda^2$ and that the curvature tensor in momentum space corresponds to a maximally symmetric space, i.e.
\begin{equation}
S_{\rho\sigma\mu\nu}\,\propto \, g_{\rho\mu}g_{\sigma\nu}-g_{\rho\nu}g_{\sigma\mu}\,.
\end{equation}

\subsection{Velocities for photons}
In this subsection we compute the velocity of photons first from an action. We start from the action 
\begin{equation}
S\,=\,\int{\left(\dot{x}^\mu k_\mu-\mathcal{N} C(x,k)\right) }d\tau
\label{eq:action1}
\end{equation} 
with the deformed Casimir of the bicrossproduct basis~\cite{KowalskiGlikman:2002we} depending of $x$ and $k$
\begin{equation}
C(\bar{k})\,=\,\Lambda^2\left(e^{\bar{k}_0/\Lambda}+e^{-\bar{k}_0/\Lambda}-2\right)- \vec{\bar{k}}^2e^{\bar{k}_0/\Lambda}\,=\,\Lambda^2\left(e^{k_0/\Lambda}+e^{-k_0/\Lambda}-2\right)-\frac{ \vec{k}^2e^{k_0/\Lambda}}{R^2(x^0)} \,.
\label{eq:Casimir_RW}
\end{equation} 
Setting $\dot{x}^0=1$, i.e. taking that the temporal coordinate as the proper time, we can obtain the value of $\mathcal{N}$ as a function of position and momenta, and then, we can obtain the velocity for massless particles (in 1+1 dimensions) as 
\begin{equation}
v\,=\,\dot{x}^1\,=\,-\frac{4 \Lambda ^3 k_1 e^{2 k_0/\Lambda} \left(e^{k_0/\Lambda}-1\right) R(x^0)^2}
{\left(k_1^2 e^{2 k_0/\Lambda}-\Lambda ^2 e^{2k_0/\Lambda} R(x^0)^2+\Lambda ^2
   R(x^0)^2\right)^2}\,.
\label{eq:velocity}
\end{equation} 
When one uses the Casimir in order to obtain $|k|$ as a function of $k_0$, one finds
\begin{equation}
k_1\,= -\,\Lambda  e^{-k_0/\Lambda}
   \left(e^{k_0/\Lambda }-1\right) R(x^0)\,,
\label{eq:RW_k}
\end{equation} 
and then, by substitution of Eq.\eqref{eq:RW_k} in Eq.\eqref{eq:velocity}, one can see that the velocity is 
\begin{equation}
v\,=\, \frac{e^{k_0/\Lambda}}{R(x^0)}\,,
\label{eq:velocity_RW_casimir}
\end{equation} 
so we will see an energy dependent velocity in this momentum coordinates. When $\Lambda$ goes to infinity one gets $v=1/R(x_0)$, which is the standard result of GR.  

This can be also obtained directly form the metric asking the line element to be null,  
\begin{equation}
0\,=\, (dx^0)^2-R(x^0)e^{-2 k_0/\Lambda}(dx^1)^2\,,
\end{equation}
which is consistent with what we claim in the previous section: the same result must be obtained starting from the action and from the line element of the metric.

\subsection{Momenta for photons}
Looking for the extrema of the action~\eqref{eq:action1}, one can find 
\begin{equation}
\dot{k}_0\,=\, - \frac{\Lambda\left(e^{k_0/\Lambda}-1\right)R'(x^0)}{R(x^0)}\,, \qquad  \dot{k}_1\,=\,0\,.
\label{eq:momenta_RW}
\end{equation}
Solving the first equation we obtain the expression of the energy as a function of time
\begin{equation}
k_0\,=\, -\Lambda \log\left(1+\frac{e^{-E/\Lambda}-1}{R(x^0)}\right) \,,
\label{eq:energy_RW}
\end{equation} 
where the constant of integration of the previous differential equation has been chosen in order to, when one takes the limit $\Lambda$ going to infinity, one recovers that the conserved energy is the bared momentum $E=k_0 R(x^0)$, so this constant can be considered as the energy conserved along the geodesic.

\subsection{Redshift}
Starting from the line element derived for photons from the metric
\begin{equation}
0\,=\,(dx^0)^2-R^2(x^0)e^{-2k_0/\Lambda}d\vec{x}^2\,,
\end{equation} 
we find
\begin{equation}
\int{\frac{dx^0\, e^{k_0/\Lambda}}{R(x^0)}}\,=\,f(x)\,.
\label{eq:step}
\end{equation} 
Now we can write Eq.~\eqref{eq:step} as a function of $x^0$ using Eq.~\eqref{eq:energy_RW} and obtaining that the quotient in frequencies are 
\begin{equation}
\frac{\nu_0}{\nu_1}\,=\,\frac{\delta t_1}{\delta t_0}\,=\,\frac{R(t_1)\left(1+(e^{-E/\Lambda}-1)/R(t_1)\right)}{R(t_0)\left(1+(e^{-E/\Lambda}-1)/R(t_0)\right)}\,=\,\frac{R(t_1)+e^{-E/\Lambda}-1}{R(t_0)+e^{-E/\Lambda}-1}\,,
\end{equation} 
and then, the redshift is 
\begin{equation}
z\,=\,\frac{R(t_0)+e^{-E/\Lambda}-1}{R(t_1)+e^{-E/\Lambda}-1}-1\,.
\label{eq:redshift}
\end{equation} 
We see that taking the limit $\Lambda\rightarrow \infty$ in the previous equation, we recover the usual redshift in Friedmann-Robertson-Walker space~\cite{Weinberg:1972kfs}. From this equation, one can observe that the redshift will be different for particles with different energies.  We can check this through a simple calculation: suppose two particles emitted from a distance source, one with energy $E\to 0$ while the other has an energy $E$, being $ E\ll\Lambda$. When detected at $R(t_0)$, the redshift will be different for each one. In particular, if we make a series expansion in the high energy scale, we see
\begin{equation}
\frac{1+z(0)}{1+z(E)}\,=\, 1+\frac{E}{\Lambda}\left(\frac{1}{R(t_0)}-\frac{1}{R(t_1)}\right)\,.
\end{equation}
Then, the redshift will be different depending on the energy of the particle we are detecting. In particular, we can observe that for higher energies there is more redshift, since 
\begin{equation}
1+z(E)\,=\,(1+z(0))\left( 1-\frac{E}{\Lambda }\left(\frac{1}{R(t_0)}-\frac{1}{R(t_1)}\right)\right)\,,
\end{equation}
where the last factor is always greater than unity since, as the universe is expanding, $R(t_1)<R(t_0)$.

\subsection{Luminosity distance}
Here we will compute the luminosity distance following the same procedure as in Ref.~\cite{Weinberg:1972kfs}. We consider a circular telescope mirror of radius $b$, placed with its center at the origin and its normal along the line of sight of the radial direction to the light source. The light rays that just graze the mirror edge form a cone at the light source that, for a locally inertial coordinate system at the source, have a half-angle $|\epsilon|$ given by the relation
\begin{equation}
b\,\approx\, R(t_0) e^{-k_0/\Lambda} x |\epsilon|\,,
\end{equation}
where $b$ is expressed here as a proper distance and $x$ is the spatial coordinate at the emission of light. Then the solid angle of this cone is 
\begin{equation}
\pi |\epsilon|^2\,=\, \frac{\pi b^2}{R^2(t_0) e^{-2 k_0/\Lambda} x^2}\,,
\end{equation}
and the fraction of all isotropically emitted photons that reach the mirror is the ratio of this solid angle to $4\pi$, or 
\begin{equation}
\frac{|\epsilon|^2}{4}\,=\, \frac{ A}{4 \pi R^2(t_0) e^{-2k_0/\Lambda}x^2}\,,
\label{eq:fract}
\end{equation}
where $A$ is the proper area of the mirror
\begin{equation}
A\,=\,\pi b^2\,.
\end{equation}
However, each photon emitted with energy $h \nu_1$ will be red-shifted to energy 
\begin{equation}
h \nu_1 \frac{R(t_1)+e^{-E/\Lambda}-1}{R(t_0)+e^{-E/\Lambda}-1}\,,
\end{equation}
and photons emitted at time intervals $\delta t_1$ will arrive at time intervals 
\begin{equation}
\delta t_1\frac{R(t_1)+e^{-E/\Lambda}-1}{R(t_0)+e^{-E/\Lambda}-1}\,,
\end{equation}
where $t_1$ is the time the light leaves the source, and $t_0$ is the time the light arrives at the mirror. Thus, the total power $P$ received by the mirror is the total power emitted by the source, its absolute luminosity $L$, times a factor 
\begin{equation}
\left(\frac{R(t_1)+e^{-E/\Lambda}-1}{R(t_0)+e^{-E/\Lambda}-1}\right)^2\,,
\end{equation}
multiplied by the fraction Eq.~\eqref{eq:fract}:
\begin{equation}
P\,=\,L\,A \frac{\left(R(t_1)+e^{-E/\Lambda}-1\right)^2}{4 \pi R^2(t_0)\left(R(t_0)+e^{-E/\Lambda}-1\right)^2 e^{-2 k_0/\Lambda}x^2}\,.
\end{equation}
The apparent luminosity $l$ is the power per unit mirror area, so using Eq.~\eqref{eq:energy_RW} we obtain
\begin{equation}
l\,\equiv\,\frac{P}{A}\,=\, L \frac{\left(R(t_1)+e^{-E/\Lambda}-1\right)^2}{4 \pi \left(R(t_0)+e^{-E/\Lambda}-1\right)^4 x^2}\,.
\label{eq:app_luminosity}
\end{equation}
In an Euclidean space the apparent luminosity of a source at rest at distance $d$ would be $L/4\pi d^2$, so in general we may define the luminosity distance $d_L$ of a light source as 
\begin{equation}
d_L\,=\,\left(\frac{L}{4\pi l}\right)^{1/2}\,,
\end{equation}
and then Eq.~\eqref{eq:app_luminosity} may therefore be written
\begin{equation} 
d_L\,=\, \frac{\left(R(t_0)+e^{-E/\Lambda}-1\right)^2\,x}{R(t_1)+e^{-E/\Lambda}-1}\,.
\end{equation}
We can rewrite the previous expression as 
\begin{equation} 
d_L\,=\,\left( \frac{R(t_0)+e^{-E/\Lambda}-1}{R(t_1)+e^{-E/\Lambda}-1}\right)^2 r\,,
\end{equation}
where 
\begin{equation} 
r\,=\,\left(R(t_1)+e^{-E/\Lambda}-1\right)x\,,
\end{equation}
is the proper distance that separates the source from us. Now we can express the luminosity distance as a function of the redshift we have found above 
 \begin{equation} 
d_L\,=\,\left(1+z\right)^2 r\,,
\end{equation}
which is the same expression one finds in GR. 

As we did for the redshift, we can see that for particles with different energies the luminosity distance will be different. One can easily find
\begin{equation}
\frac{d_L (0)}{d_L (E)}\,=\,\left(\frac{1+z(0)}{1+z(E)}\right)^2\,,
\end{equation}
and then, the luminosity distance will be an increasing function of energy, as the redshift is. This is an interesting feature that perhaps could be tested in the future in cosmographic analyses.   

\subsection{Congruence of geodesics}
In this part we will study the congruence of null geodesics for the metric of the cotangent bundle. We make the computation taking the procedure of Ref.~\cite{Poisson:2009pwt}. 

We start from the definition of the expansion for null geodesics 
\begin{equation}
\theta \,=\,\frac{1}{\delta S}\frac{d }{d \lambda}\delta S\,,
\end{equation}
where $\delta S$ is the infinitesimal change of surface. For the metric of Eq.~\eqref{eq:RW_metric} we obtain
\begin{equation}
\theta\,=\,2\frac{e^{k_0/\Lambda}R^\prime(t)}{R^2(t)}\,,
\label{eq:theta_RW}
\end{equation}
where $R^\prime(t)=d R(t)/dt$. Taking this expression we can see that making a series expansion in $\theta$ 
\begin{equation}
\frac{\theta(0)}{\theta(E)} \,=\, 1-\frac{E}{R(t) \Lambda }\,.
\end{equation}
The expansion of the congruence of the geodesics will depend on the energy, in such a way that the expansion will be greater for larger energies, since 
\begin{equation}
\theta(E_h) \,=\,\theta(E_l)\left( 1+\frac{E}{R(t) \Lambda }\right)\,.
\end{equation}

\section{Schwarzschild metric}
\label{sec:sch}

Now we will focus on the Schwarzschild solution. We choose the tetrad of Lemaître coordinates~\cite{Landau:1982dva}\footnote{The use of Lemaitre coordinates is necessary because in the most common choices of  coordinates~\cite{Poisson:2009pwt}, the metric is singular in the horizon because of the momentum dependent term.}
\begin{equation}
e^t_t\,=\,1\,,\qquad e^x_x\,=\, \sqrt{\frac{2M}{r}}\,,\qquad e^\theta_\theta(x)\,=\, r\,,\qquad e^\phi_\phi(x)\,=\, r \sin{\theta}\,,
\label{eq:Sch_tetrad}
\end{equation}
where 
\begin{equation}
r\,=\,\left(\frac{3}{2}\left(x-t\right)\right)^{(2/3)}\left(2M\right)^{(1/3)}\,.
\end{equation}
With the same choice of the momentum tetrad of Sec.~\ref{sec:rw}, we obtain from  Eq.~\eqref{eq:cotangent_metric_tetrads} the metric in the cotangent bundle
\begin{equation}
\begin{split}
g_{tt}(x,k)\,&=\,1\,, \qquad  g_{xx}(x,k)\,=\,-\frac{2M}{r}e^{-2 k_0/\Lambda}\,, \\
g_{\theta\theta}(x,k)\,&=\, -r^2e^{-2 k_0/\Lambda}\,, \qquad  g_{\phi\phi}(x,k)\,=\,- r^2 \sin^2{\theta}e^{-2 k_0/\Lambda}\,.
\label{eq:Sch_metric}
\end{split}
\end{equation}
Again, one can see that the momentum scalar of curvature is constant $S=12/\Lambda^2$ and that the momentum curvature tensor corresponds to a maximally symmetric space. 

Now we will study the event horizon in this modified metric of Schwarzschild. First, we will see the conserved energy along geodesics. After that, we will represent the null geodesics in order to obtain the event horizon. Finally, we will compute the surface gravity. 

\subsection{Conserved energy} 
In the case we are considering, Eq.\eqref{eq:killing} gives 
\begin{equation}
\chi^0\,=\,1\,,\qquad \chi^1\,=\,1\,,
\end{equation}
which is exactly the same Killing vector obtained in GR~\footnote{This can be easily understood just looking at Eq.\eqref{eq:killing}. If in GR there is a constant Killing vector, the same vector will be a Killing one in this modified equation.}.

One also can get the same result from the action Eq.~\eqref{eq:action1} where the Casimir is 
\begin{equation}
C(\bar{k})\,=\,\Lambda^2\left(e^{\bar{k}_0/\Lambda}+e^{-\bar{k}_0/\Lambda}-2\right)- \vec{\bar{k}}^2e^{\bar{k}_0/\Lambda}\,=\,\Lambda^2\left(e^{k_0/\Lambda}+e^{-k_0/\Lambda}-2\right)- \vec{k}^2e^{k_0/\Lambda}\frac{r}{2M} \,.
\label{eq:Casimir_Sch}
\end{equation} 
With the choice of $\tau=t$, one can express $\mathcal{N}$ in Eq.~\eqref{eq:action1} as a function of $x$ and $k$ and then, it is easy to see that the derivatives of the momenta satisfy (in 1+1 dimensions)
\begin{equation}
\dot{k}_0+\dot{k}_1\,=\,0\,.
\end{equation} 
Using the Casimir, we find for photons the relation between the spatial component of the momentum and the zero component
\begin{equation}
k_1\,=\,\sqrt{\frac{2M}{r}}\Lambda \left(1-e^{-k_0/\Lambda}\right)\,,
\end{equation} 
so the conserved energy is 
\begin{equation}
E\,=\,k_0+k_1\,=\,k_0+\sqrt{\frac{2M}{r}}\Lambda \left(1-e^{-k_0/\Lambda}\right)\,.
\label{eq:energy_Sch}
\end{equation} 

\subsection{Event horizon} 
\begin{figure}[H]
  \includegraphics[width=\linewidth]{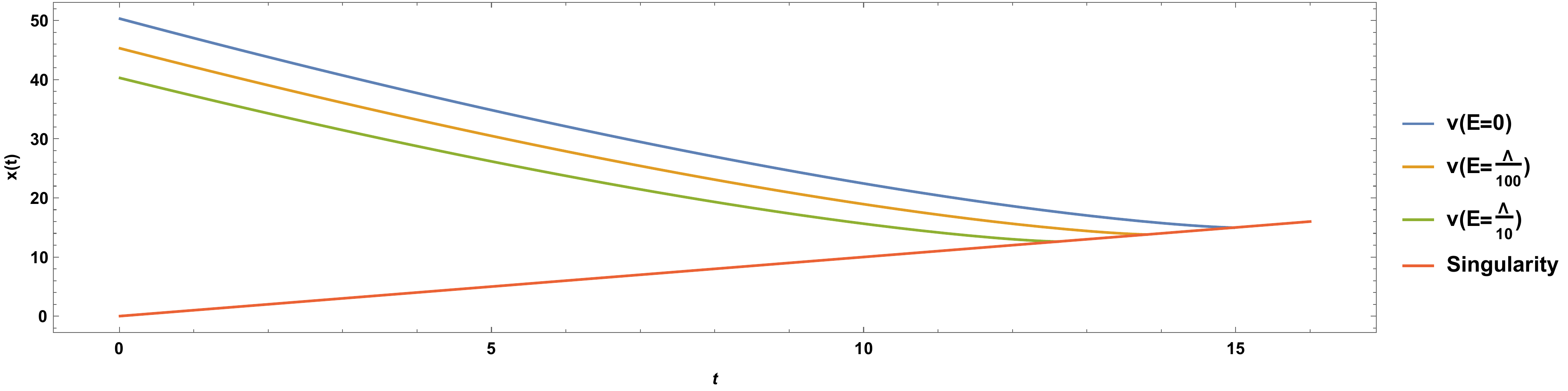}
  \caption{Particles  with three different velocities coming from outside the horizon, crossing it and finally arriving to the singularity.}
  \label{fig:ingoing}
\end{figure}

\begin{figure}[H]
  \includegraphics[width=\linewidth]{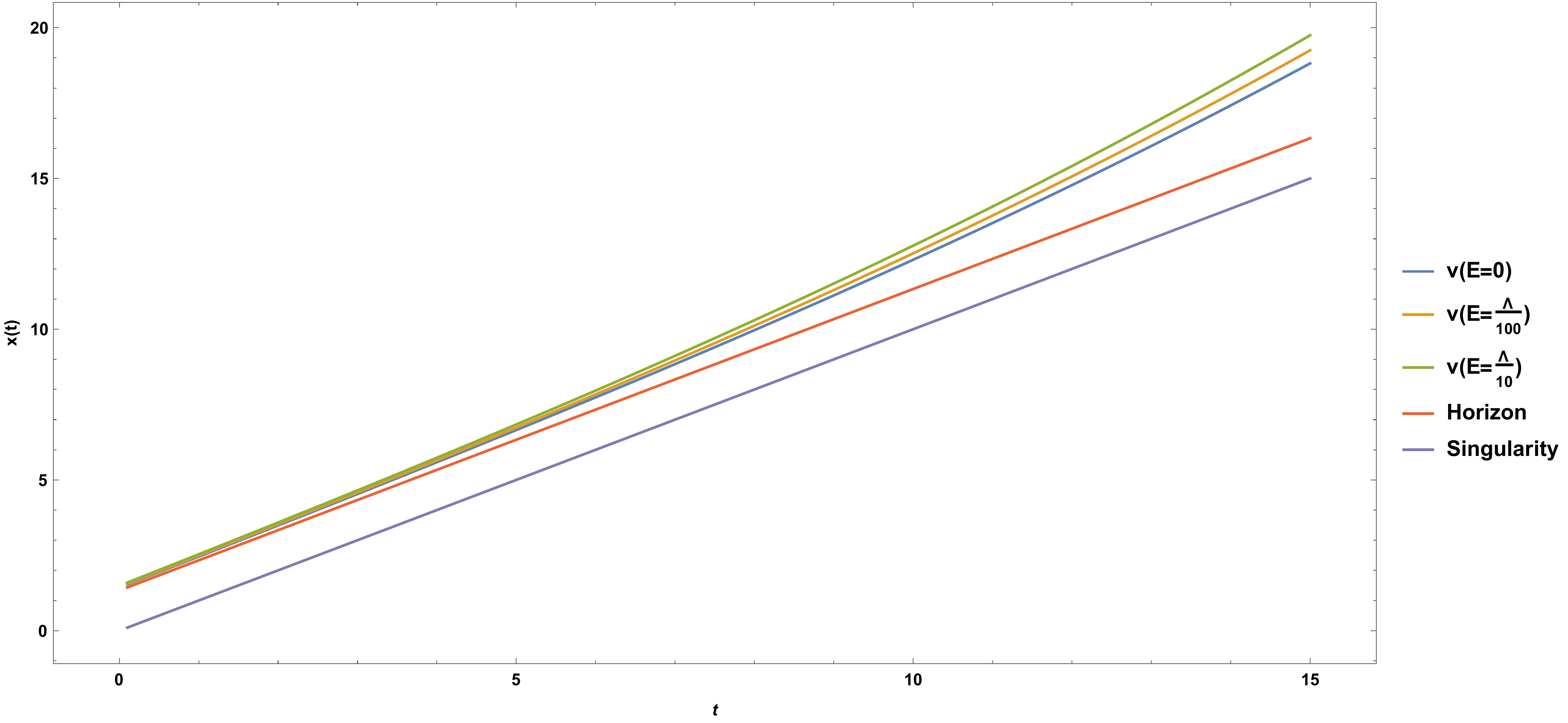}
  \caption{Outgoing null geodesics from outside the horizon.}
  \label{fig:outgoing}
\end{figure}

\begin{figure}[H]
  \includegraphics[width=\linewidth]{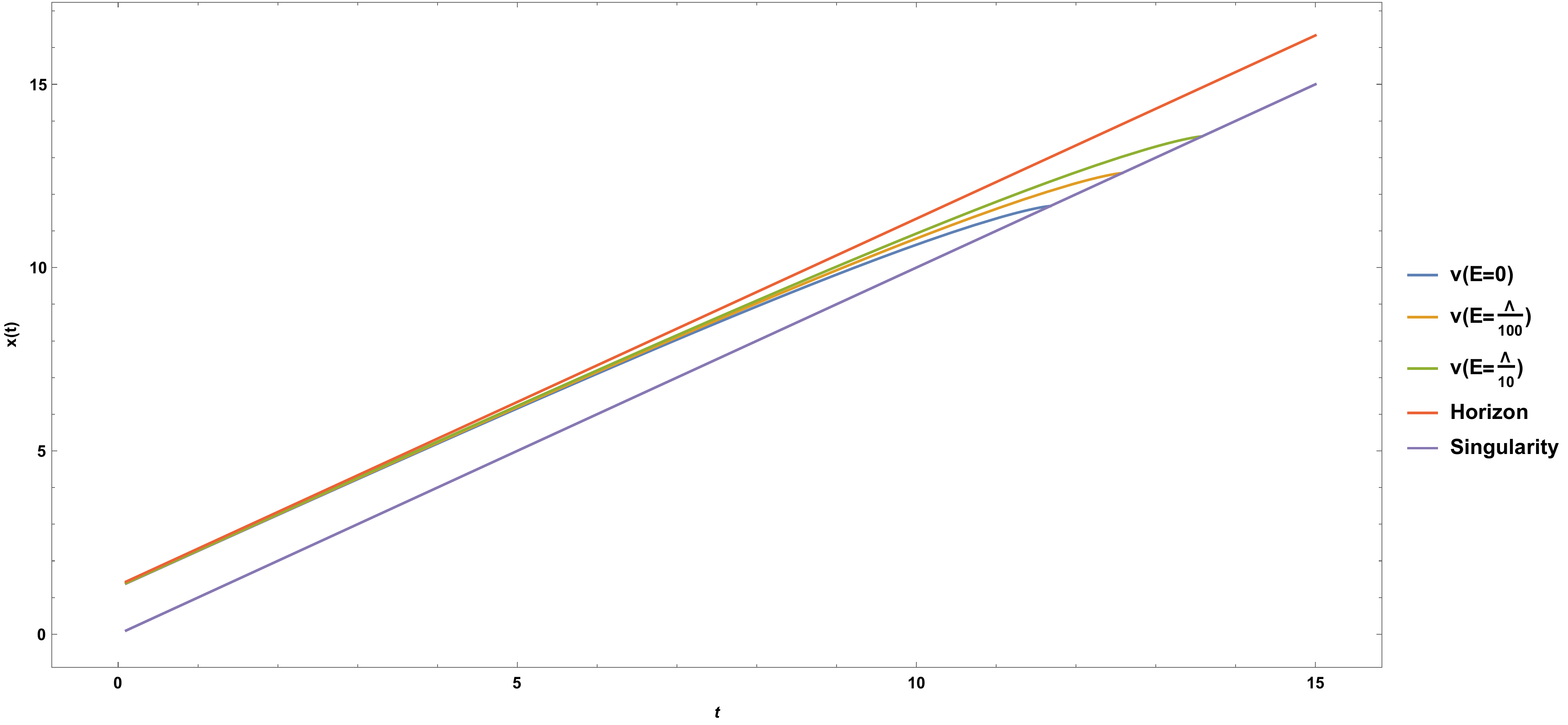}
  \caption{Null geodesics from inside the horizon falling at the singularity.}
  \label{fig:outgoing_2}
\end{figure}

In order to compute the event horizon, we study the null ingoing and outgoing geodesics. In GR, the horizon in the these coordinates is in $x-t=4 M/3$ and the singularity is at  $x=t$~\cite{Landau:1982dva}. We first start from the line element of the metric Eq.~\eqref{eq:Sch_metric}. Then we can solve the differential equation 
\begin{equation}
ds^2\,=\,0\,\implies\,\frac{dx}{dt}\,=\,\pm \left(\frac{3(x-t)}{4 M}\right)^{(1/3)}e^{k_0/\Lambda}\,,
\label{eq:eq_geo}
\end{equation}
where + stands for outgoing geodesics and - for ingoing. We can solve numerically this differential equation writing $k_0$ as a function of the conserved energy (inverting Eq.~\eqref{eq:energy_Sch}) and then, plot it  for different energies. We observe that doing the numerical computation one sees no difference between the geodesics with different energies. Taking $M=1$ we show the behavior for ingoing geodesics\footnote{In the next figures, the representation of the geodesics are carried out for different initial conditions for different energies, and hence the trajectories are different.} in Fig.~\ref{fig:ingoing}. 
For particles emitted outside the horizon but close to it, they will escape in a finite time, see Fig.~\ref{fig:outgoing}.
Also, we can represent the geodesics starting inside the horizon in Fig.~\ref{fig:outgoing_2}.
and we see that they finally go into the singularity. In Refs.~\cite{Dubovsky:2006vk,Barausse:2011pu,Blas:2011ni,Bhattacharyya:2015gwa} it is showed that in a LIV scenario, where there are different horizons for particles with different energies, there is a violation of the second law of the black hole thermodynamics making possible a construction of a perpetuum mobile (see however~\cite{Benkel:2018abt} for a possible resolution of this problem). We see that in contrast with LIV scenarios, in our case there is a unique horizon, which is consistent with the fact that in DSR framework there is a relativity principle.

\subsection{Surface gravity}
There are different ways to compute the surface gravity of a black hole~\cite{Cropp:2013zxi}. In particular, it can be related with the peeling off properties of null geodesics near the horizon. In~\cite{Cropp:2013zxi} it was shown that the surface gravity can be defined as
\begin{equation}
\frac{d |x_1(t)-x_2(t)| }{d t}\,\approx \,  \kappa_{\text{peeling}}(t) |x_1(t)-x_2(t)|\,,
\label{eq:surface_gravity}
\end{equation}
where $x_1(t)$ and $x_2(t)$ are two null geodesics on the same side of the horizon and the normalization of $\kappa_{\text{peeling}}$ is chosen so to coincide with $\kappa_{\text{inaffinity}}$ in the GR limit. We obtain from Eq.~\eqref{eq:eq_geo}, for two null geodesics with the same energy $k_0$:
\begin{equation}
\frac{d |x_1(t)-x_2(t)| }{d t}\,\approx \,\frac{e^{k_0/\Lambda}}{4M} |x_1(t)-x_2(t)|\,,
\end{equation}
and then, 
\begin{equation}
 \kappa_{\text{peeling}}\,=\,\frac{e^{k_0/\Lambda}}{4M}\,,
\end{equation}
which depends on the energy of the geodesic. This seems to imply that the Hawking temperature defined as~\cite{Poisson:2009pwt}
\begin{equation}
T\,=\,\frac{\kappa}{2 \pi}\,,
\end{equation}
will generally depend on the energy of the outgoing particles, an interesting result that deserves further investigation and we leave for future work.

\section{Conclusions}
\label{sec:conclusions}
In this paper we proposed a generalization of a curved spacetime that takes into account a curvature of momentum space. The obtained  metric is invariant under space-time diffeomorphisms and hence, the results we find are independent of the space-time coordinates one uses as in GR, but depend on the coordinates of momentum space. We have shown that if one considers the dispersion relation as the squared distance in momentum space, the study of propagation of a particle through a metric or a phase-space action leads to the same results.  In particular, we have considered a de Sitter momentum space, which represents the modified kinematics of $\kappa$-Poincaré, and we considered case studies: Friedmann-Robertson-Walker universe and the Schwarzschild black hole.  

In the Friedmann-Robertson-Walker metric we have studied the modified geodesics, redshift, luminosity distance and the expansion of geodesics. We saw that higher energetic photons have greater velocity than the lower energetic ones implying for them greater redshift, luminosity distance and geodesic expansion (this depends on the sign appearing with the high energy scale $\Lambda$, varying the results if the sign changes).

For the Schwarzschild metric, we have studied the null geodesics showing that particles with different energies still have the same horizon, in contrast with the LIV case~\cite{Kifune:1999ex,Dubovsky:2006vk}, where there are different horizons for particles with different energies. This is in agreement with the preserved relativity principle of DSR. However, the surface gravity computed from the peeling off of null geodesics is energy dependent, suggesting that the Hawking's temperature will depend on the energy.  All of these phenomena could be used to constrain the scale $\Lambda$ within this framework. We hope to explore these implications in future works. 

\section*{Acknowledgments}
This work is supported by the Spanish grants PGC2018-095328-B-I00 (FEDER/Agencia estatal de investigación), and by the Spanish DGIID-DGA Grant No. 2015-E24/2.
The authors would also like to thank support from the COST Action CA18108. We acknowledge useful discussions with César Asensio, Jesús Clemente, José Manuel Carmona, José Luis Cortés and specially Raúl Carballo-Rubio.

\appendix
\section{Scalar of curvature of the momentum space}
\label{sec:appendix}

In this appendix we show that, when one considers a metric in the cotangent bundle constructed from a metric in momentum space of constant curvature,  the scalar of curvature of the momentum space is also constant. We start by the definition of the curvature tensor in momentum space of Eq.~\eqref{eq:Riemann_p} for flat spacetime
\begin{equation}
S_{\sigma}^{\mu\nu\rho}(k)\,=\, \frac{\partial C^{\mu\nu}_\sigma(k)}{\partial k_\rho}-\frac{\partial C^{\mu\rho}_\sigma(k)}{\partial k_\nu}+C_\sigma^{\lambda\nu}(k) \,C^{\mu\rho}_\lambda(k)-C_\sigma^{\lambda\rho}(k)\,C^{\mu\nu}_\lambda(k)\,,
\end{equation} 
 that can be rewritten using Eq.~\eqref{eq:affine_connection_p} and raising the low index as 
\begin{equation}
S^{\sigma\kappa\lambda\mu}(k)\,=\, \frac{1}{2}\left(\frac{\partial^2 g_k^{\sigma\mu}(k)}{\partial k_\kappa \partial k_\lambda}+\frac{\partial^2 g_k^{\kappa\lambda}(k)}{\partial k_\sigma \partial k_\mu}-\frac{\partial^2 g_k^{\sigma\lambda}(k)}{\partial k_\kappa \partial k_\mu}-\frac{\partial^2 g_k^{\kappa\mu}(k)}{\partial k_\sigma \partial k_\lambda}\right)
+g_k^{\nu\tau}(k)\left(C_\nu^{\kappa\lambda}(k)\,C^{\sigma\mu}_\tau(k)-C_\nu^{\kappa\mu}(k)\,C^{\sigma\lambda}_\tau(k)\right)\,.
\end{equation} 
Our principal assumption claims that we must do the change $k\rightarrow \bar{k}=\bar{e}k$ in order to consider a nontrivial geometry in spacetime, so the previous equation should be 
\begin{equation}
S^{\sigma\kappa\lambda\mu}(\bar{k})\,=\, \frac{1}{2}\left(\frac{\partial^2 g_{\bar{k}}^{\sigma\mu}(\bar{k})}{\partial \bar{k}_\kappa \partial \bar{k}_\lambda}+\frac{\partial^2 g_{\bar{k}}^{\kappa\lambda}(\bar{k})}{\partial \bar{k}_\sigma \partial \bar{k}_\mu}-\frac{\partial^2 g_{\bar{k}}^{\sigma\lambda}(\bar{k})}{\partial \bar{k}_\kappa \partial \bar{k}_\mu}-\frac{\partial^2 g_{\bar{k}}^{\kappa\mu}(\bar{k})}{\partial \bar{k}_\sigma \partial \bar{k}_\lambda}\right)
+g_{\bar{k}}^{\nu\tau}(\bar{k})\left(C_\nu^{\kappa\lambda}(\bar{k})\,C^{\sigma\mu}_\tau(\bar{k})-C_\nu^{\kappa\mu}(\bar{k})\,C^{\sigma\lambda}_\tau(\bar{k})\right)\,,
\end{equation}
which contracting gives 
\begin{equation}
S^{\sigma\kappa\lambda\mu}(\bar{k})g^{\bar{k}}_{\sigma\lambda}(\bar{k})g^{\bar{k}}_{\kappa\mu}(\bar{k})\,=\,\text{const}\,,
\end{equation}
since the momentum space is maximally symmetric, and where $g^{\bar{k}}_{\kappa\nu}(\bar{k})$ is the inverse of the metric 
\begin{equation}
g^{\bar{k}}_{\kappa\nu}(\bar{k})g_{\bar{k}}^{\kappa\mu}(\bar{k})\,=\,\delta^\mu_\nu\,.
\end{equation}
Now we can compute the scalar of curvature in momentum space from the metric in the cotangent bundle given by
\begin{equation}
g_{\mu\nu}(x,k)\,=\,e^\rho_\mu(x)g^{\bar{k}}_{\rho\sigma}(\bar{k})e^\sigma_\nu(x)\,,
\end{equation}
and where the curvature tensor in momentum space is now
\begin{equation}
\begin{split}
S^{\sigma\kappa\lambda\mu}(x,k)\,&=\, \frac{1}{2}\left(\frac{\partial^2 g^{\sigma\mu}(x,k)}{\partial k_\kappa \partial k_\lambda}+\frac{\partial^2 g^{\kappa\lambda}(x,k)}{\partial k_\sigma \partial k_\mu}-\frac{\partial^2 g^{\sigma\lambda}(x,k)}{\partial k_\kappa \partial k_\mu}-\frac{\partial^2 g^{\kappa\mu}(x,k)}{\partial k_\sigma \partial k_\lambda}\right)\\
&+g^{\nu\tau}(x,k)\left(C_\nu^{\kappa\lambda}(x,k)\,C^{\sigma\mu}_\tau(x,k)-C_\nu^{\kappa\mu}(x,k)\,C^{\sigma\lambda}_\tau(x,k)\right)\,.
\end{split}
\end{equation} 
After some steps, one can finally check that 
\begin{equation}
S^{\sigma\kappa\lambda\mu}(x,k)g_{\sigma\lambda}(x,k)g_{\kappa\mu}(x,k)\,=\,S^{\sigma\kappa\lambda\mu}(\bar{k})g^{\bar{k}}_{\sigma\lambda}(\bar{k})g^{\bar{k}}_{\kappa\mu}(\bar{k})\,=\,\text{const}\,,
\end{equation}
so with our procedure we still have a constant curvature momentum space and therefore, it is not strange that we have found ten transformations for momenta for a fixed point $x$ (that we can call momentum isometries of the metric): four of them are related with translations and the other six are the transformations that leave the momentum origin invariant (i.e. the point in phase space $(x,0)$).


\begin{thebibliography}{50}%
\makeatletter
\providecommand \@ifxundefined [1]{%
 \@ifx{#1\undefined}
}%
\providecommand \@ifnum [1]{%
 \ifnum #1\expandafter \@firstoftwo
 \else \expandafter \@secondoftwo
 \fi
}%
\providecommand \@ifx [1]{%
 \ifx #1\expandafter \@firstoftwo
 \else \expandafter \@secondoftwo
 \fi
}%
\providecommand \natexlab [1]{#1}%
\providecommand \enquote  [1]{``#1''}%
\providecommand \bibnamefont  [1]{#1}%
\providecommand \bibfnamefont [1]{#1}%
\providecommand \citenamefont [1]{#1}%
\providecommand \href@noop [0]{\@secondoftwo}%
\providecommand \href [0]{\begingroup \@sanitize@url \@href}%
\providecommand \@href[1]{\@@startlink{#1}\@@href}%
\providecommand \@@href[1]{\endgroup#1\@@endlink}%
\providecommand \@sanitize@url [0]{\catcode `\\12\catcode `\$12\catcode
  `\&12\catcode `\#12\catcode `\^12\catcode `\_12\catcode `\%12\relax}%
\providecommand \@@startlink[1]{}%
\providecommand \@@endlink[0]{}%
\providecommand \url  [0]{\begingroup\@sanitize@url \@url }%
\providecommand \@url [1]{\endgroup\@href {#1}{\urlprefix }}%
\providecommand \urlprefix  [0]{URL }%
\providecommand \Eprint [0]{\href }%
\providecommand \doibase [0]{http://dx.doi.org/}%
\providecommand \selectlanguage [0]{\@gobble}%
\providecommand \bibinfo  [0]{\@secondoftwo}%
\providecommand \bibfield  [0]{\@secondoftwo}%
\providecommand \translation [1]{[#1]}%
\providecommand \BibitemOpen [0]{}%
\providecommand \bibitemStop [0]{}%
\providecommand \bibitemNoStop [0]{.\EOS\space}%
\providecommand \EOS [0]{\spacefactor3000\relax}%
\providecommand \BibitemShut  [1]{\csname bibitem#1\endcsname}%
\let\auto@bib@innerbib\@empty
\bibitem [{\citenamefont {Mukhi}(2011)}]{Mukhi:2011zz}%
  \BibitemOpen
  \bibfield  {author} {\bibinfo {author} {\bibfnamefont {S.}~\bibnamefont
  {Mukhi}},\ }\href {\doibase 10.1088/0264-9381/28/15/153001} {\bibfield
  {journal} {\bibinfo  {journal} {Class. Quant. Grav.}\ }\textbf {\bibinfo
  {volume} {28}},\ \bibinfo {pages} {153001} (\bibinfo {year} {2011})},\
  \Eprint {http://arxiv.org/abs/1110.2569} {arXiv:1110.2569 [physics.pop-ph]}
  \BibitemShut {NoStop}%
\bibitem [{\citenamefont {Aharony}(2000)}]{Aharony:1999ks}%
  \BibitemOpen
  \bibfield  {author} {\bibinfo {author} {\bibfnamefont {O.}~\bibnamefont
  {Aharony}},\ }\bibfield  {booktitle} {\emph {\bibinfo {booktitle} {{Strings
  '99. Proceedings, Conference, Potsdam, Germany, July 19-24, 1999}}},\ }\href
  {\doibase 10.1088/0264-9381/17/5/302} {\bibfield  {journal} {\bibinfo
  {journal} {Class. Quant. Grav.}\ }\textbf {\bibinfo {volume} {17}},\ \bibinfo
  {pages} {929} (\bibinfo {year} {2000})},\ \Eprint
  {http://arxiv.org/abs/hep-th/9911147} {arXiv:hep-th/9911147 [hep-th]}
  \BibitemShut {NoStop}%
\bibitem [{\citenamefont {Dienes}(1997)}]{Dienes:1996du}%
  \BibitemOpen
  \bibfield  {author} {\bibinfo {author} {\bibfnamefont {K.~R.}\ \bibnamefont
  {Dienes}},\ }\bibfield  {booktitle} {\emph {\bibinfo {booktitle} {{Institute
  for Theoretical Physics Conference on Unification: From the Weak Scale to the
  Planck Scale Santa Barbara, California, October 23-27, 1995}}},\ }\href
  {\doibase 10.1016/S0370-1573(97)00009-4} {\bibfield  {journal} {\bibinfo
  {journal} {Phys. Rept.}\ }\textbf {\bibinfo {volume} {287}},\ \bibinfo
  {pages} {447} (\bibinfo {year} {1997})},\ \Eprint
  {http://arxiv.org/abs/hep-th/9602045} {arXiv:hep-th/9602045 [hep-th]}
  \BibitemShut {NoStop}%
\bibitem [{\citenamefont {Sahlmann}(2010)}]{Sahlmann:2010zf}%
  \BibitemOpen
  \bibfield  {author} {\bibinfo {author} {\bibfnamefont {H.}~\bibnamefont
  {Sahlmann}},\ }in\ \href
  {https://inspirehep.net/record/843661/files/arXiv:1001.4188.pdf} {\emph
  {\bibinfo {booktitle} {{Proceedings, Foundations of Space and Time:
  Reflections on Quantum Gravity: Cape Town, South Africa}}}}\ (\bibinfo {year}
  {2010})\ pp.\ \bibinfo {pages} {185--210},\ \Eprint
  {http://arxiv.org/abs/1001.4188} {arXiv:1001.4188 [gr-qc]} \BibitemShut
  {NoStop}%
\bibitem [{\citenamefont {Dupuis}\ \emph {et~al.}(2012)\citenamefont {Dupuis},
  \citenamefont {Ryan},\ and\ \citenamefont {Speziale}}]{Dupuis:2012yw}%
  \BibitemOpen
  \bibfield  {author} {\bibinfo {author} {\bibfnamefont {M.}~\bibnamefont
  {Dupuis}}, \bibinfo {author} {\bibfnamefont {J.~P.}\ \bibnamefont {Ryan}}, \
  and\ \bibinfo {author} {\bibfnamefont {S.}~\bibnamefont {Speziale}},\ }\href
  {\doibase 10.3842/SIGMA.2012.052} {\bibfield  {journal} {\bibinfo  {journal}
  {SIGMA}\ }\textbf {\bibinfo {volume} {8}},\ \bibinfo {pages} {052} (\bibinfo
  {year} {2012})},\ \Eprint {http://arxiv.org/abs/1204.5394} {arXiv:1204.5394
  [gr-qc]} \BibitemShut {NoStop}%
\bibitem [{\citenamefont {Van~Nieuwenhuizen}(1981)}]{VanNieuwenhuizen:1981ae}%
  \BibitemOpen
  \bibfield  {author} {\bibinfo {author} {\bibfnamefont {P.}~\bibnamefont
  {Van~Nieuwenhuizen}},\ }\href {\doibase 10.1016/0370-1573(81)90157-5}
  {\bibfield  {journal} {\bibinfo  {journal} {Phys. Rept.}\ }\textbf {\bibinfo
  {volume} {68}},\ \bibinfo {pages} {189} (\bibinfo {year} {1981})}\BibitemShut
  {NoStop}%
\bibitem [{\citenamefont {Taylor}(1984)}]{Taylor:1983su}%
  \BibitemOpen
  \bibfield  {author} {\bibinfo {author} {\bibfnamefont {J.~G.}\ \bibnamefont
  {Taylor}},\ }\href {\doibase 10.1016/0146-6410(84)90002-4} {\bibfield
  {journal} {\bibinfo  {journal} {Prog. Part. Nucl. Phys.}\ }\textbf {\bibinfo
  {volume} {12}},\ \bibinfo {pages} {1} (\bibinfo {year} {1984})}\BibitemShut
  {NoStop}%
\bibitem [{\citenamefont {Wallden}(2013)}]{Wallden:2013kka}%
  \BibitemOpen
  \bibfield  {author} {\bibinfo {author} {\bibfnamefont {P.}~\bibnamefont
  {Wallden}},\ }\bibfield  {booktitle} {\emph {\bibinfo {booktitle}
  {{Proceedings, 15th Conference on Recent Developments in Gravity (NEB 15):
  Chania, Crete, Greece, June 20-23, 2012}}},\ }\href {\doibase
  10.1088/1742-6596/453/1/012023} {\bibfield  {journal} {\bibinfo  {journal}
  {J. Phys. Conf. Ser.}\ }\textbf {\bibinfo {volume} {453}},\ \bibinfo {pages}
  {012023} (\bibinfo {year} {2013})}\BibitemShut {NoStop}%
\bibitem [{\citenamefont {Wallden}(2010)}]{Wallden:2010sh}%
  \BibitemOpen
  \bibfield  {author} {\bibinfo {author} {\bibfnamefont {P.}~\bibnamefont
  {Wallden}},\ }\bibfield  {booktitle} {\emph {\bibinfo {booktitle} {{Classical
  and quantum gravity. Proceedings, 1st Mediterranean Conference, MCCQG 2009,
  Kolymbari, Crete, Greece, September 14-18, 2009}}},\ }\href {\doibase
  10.1088/1742-6596/222/1/012053} {\bibfield  {journal} {\bibinfo  {journal}
  {J. Phys. Conf. Ser.}\ }\textbf {\bibinfo {volume} {222}},\ \bibinfo {pages}
  {012053} (\bibinfo {year} {2010})},\ \Eprint {http://arxiv.org/abs/1001.4041}
  {arXiv:1001.4041 [gr-qc]} \BibitemShut {NoStop}%
\bibitem [{\citenamefont {Henson}(2009)}]{Henson:2006kf}%
  \BibitemOpen
  \bibfield  {author} {\bibinfo {author} {\bibfnamefont {J.}~\bibnamefont
  {Henson}},\ }in\ \href@noop {} {\emph {\bibinfo {booktitle} {Approaches to
  Quantum Gravity: Toward a New Understanding of Space, Time and Matter}}},\
  \bibinfo {editor} {edited by\ \bibinfo {editor} {\bibfnamefont
  {D.}~\bibnamefont {Oriti}}}\ (\bibinfo  {publisher} {Cambridge University
  Press},\ \bibinfo {year} {2009})\ pp.\ \bibinfo {pages} {393--413},\ \Eprint
  {http://arxiv.org/abs/gr-qc/0601121} {arXiv:gr-qc/0601121 [gr-qc]}
  \BibitemShut {NoStop}%
\bibitem [{\citenamefont {Gross}\ and\ \citenamefont
  {Mende}(1988)}]{Gross:1987ar}%
  \BibitemOpen
  \bibfield  {author} {\bibinfo {author} {\bibfnamefont {D.~J.}\ \bibnamefont
  {Gross}}\ and\ \bibinfo {author} {\bibfnamefont {P.~F.}\ \bibnamefont
  {Mende}},\ }\href {\doibase 10.1016/0550-3213(88)90390-2} {\bibfield
  {journal} {\bibinfo  {journal} {Nucl. Phys.}\ }\textbf {\bibinfo {volume}
  {B303}},\ \bibinfo {pages} {407} (\bibinfo {year} {1988})}\BibitemShut
  {NoStop}%
\bibitem [{\citenamefont {Amati}\ \emph {et~al.}(1989)\citenamefont {Amati},
  \citenamefont {Ciafaloni},\ and\ \citenamefont {Veneziano}}]{Amati:1988tn}%
  \BibitemOpen
  \bibfield  {author} {\bibinfo {author} {\bibfnamefont {D.}~\bibnamefont
  {Amati}}, \bibinfo {author} {\bibfnamefont {M.}~\bibnamefont {Ciafaloni}}, \
  and\ \bibinfo {author} {\bibfnamefont {G.}~\bibnamefont {Veneziano}},\ }\href
  {\doibase 10.1016/0370-2693(89)91366-X} {\bibfield  {journal} {\bibinfo
  {journal} {Phys. Lett.}\ }\textbf {\bibinfo {volume} {B216}},\ \bibinfo
  {pages} {41} (\bibinfo {year} {1989})}\BibitemShut {NoStop}%
\bibitem [{\citenamefont {Garay}(1995)}]{Garay1995}%
  \BibitemOpen
  \bibfield  {author} {\bibinfo {author} {\bibfnamefont {L.~J.}\ \bibnamefont
  {Garay}},\ }\href {\doibase 10.1142/S0217751X95000085} {\bibfield  {journal}
  {\bibinfo  {journal} {Int. J. Mod. Phys.}\ }\textbf {\bibinfo {volume}
  {A10}},\ \bibinfo {pages} {145} (\bibinfo {year} {1995})},\ \Eprint
  {http://arxiv.org/abs/gr-qc/9403008} {arXiv:gr-qc/9403008 [gr-qc]}
  \BibitemShut {NoStop}%
\bibitem [{\citenamefont {Rovelli}\ and\ \citenamefont
  {Speziale}(2003)}]{Rovelli:2002vp}%
  \BibitemOpen
  \bibfield  {author} {\bibinfo {author} {\bibfnamefont {C.}~\bibnamefont
  {Rovelli}}\ and\ \bibinfo {author} {\bibfnamefont {S.}~\bibnamefont
  {Speziale}},\ }\href {\doibase 10.1103/PhysRevD.67.064019} {\bibfield
  {journal} {\bibinfo  {journal} {Phys. Rev.}\ }\textbf {\bibinfo {volume}
  {D67}},\ \bibinfo {pages} {064019} (\bibinfo {year} {2003})},\ \Eprint
  {http://arxiv.org/abs/gr-qc/0205108} {arXiv:gr-qc/0205108 [gr-qc]}
  \BibitemShut {NoStop}%
\bibitem [{\citenamefont {Mattingly}(2005)}]{Mattingly:2005re}%
  \BibitemOpen
  \bibfield  {author} {\bibinfo {author} {\bibfnamefont {D.}~\bibnamefont
  {Mattingly}},\ }\href@noop {} {\bibfield  {journal} {\bibinfo  {journal}
  {Living Rev.Rel.}\ }\textbf {\bibinfo {volume} {8}},\ \bibinfo {pages} {5}
  (\bibinfo {year} {2005})},\ \Eprint {http://arxiv.org/abs/gr-qc/0502097}
  {arXiv:gr-qc/0502097 [gr-qc]} \BibitemShut {NoStop}%
\bibitem [{\citenamefont {Liberati}(2013)}]{Liberati2013}%
  \BibitemOpen
  \bibfield  {author} {\bibinfo {author} {\bibfnamefont {S.}~\bibnamefont
  {Liberati}},\ }\href {\doibase 10.1088/0264-9381/30/13/133001} {\bibfield
  {journal} {\bibinfo  {journal} {Class.Quant.Grav.}\ }\textbf {\bibinfo
  {volume} {30}},\ \bibinfo {pages} {133001} (\bibinfo {year} {2013})},\
  \Eprint {http://arxiv.org/abs/1304.5795} {arXiv:1304.5795 [gr-qc]}
  \BibitemShut {NoStop}%
\bibitem [{\citenamefont {Amelino-Camelia}(2013)}]{AmelinoCamelia:2008qg}%
  \BibitemOpen
  \bibfield  {author} {\bibinfo {author} {\bibfnamefont {G.}~\bibnamefont
  {Amelino-Camelia}},\ }\href {\doibase 10.12942/lrr-2013-5} {\bibfield
  {journal} {\bibinfo  {journal} {Living Rev.Rel.}\ }\textbf {\bibinfo {volume}
  {16}},\ \bibinfo {pages} {5} (\bibinfo {year} {2013})},\ \Eprint
  {http://arxiv.org/abs/0806.0339} {arXiv:0806.0339 [gr-qc]} \BibitemShut
  {NoStop}%
\bibitem [{\citenamefont {Majid}(1995)}]{Majid:1995qg}%
  \BibitemOpen
  \bibfield  {author} {\bibinfo {author} {\bibfnamefont {S.}~\bibnamefont
  {Majid}},\ }\href@noop {} {\emph {\bibinfo {title} {Foundations of Quantum
  Group Theory}}}\ (\bibinfo  {publisher} {Cambridge University Press},\
  \bibinfo {year} {1995})\BibitemShut {NoStop}%
\bibitem [{\citenamefont {Majid}\ and\ \citenamefont
  {Ruegg}(1994)}]{Majid1994}%
  \BibitemOpen
  \bibfield  {author} {\bibinfo {author} {\bibfnamefont {S.}~\bibnamefont
  {Majid}}\ and\ \bibinfo {author} {\bibfnamefont {H.}~\bibnamefont {Ruegg}},\
  }\href {\doibase 10.1016/0370-2693(94)90699-8} {\bibfield  {journal}
  {\bibinfo  {journal} {Phys. Lett.}\ }\textbf {\bibinfo {volume} {B334}},\
  \bibinfo {pages} {348} (\bibinfo {year} {1994})},\ \Eprint
  {http://arxiv.org/abs/hep-th/9405107} {arXiv:hep-th/9405107 [hep-th]}
  \BibitemShut {NoStop}%
\bibitem [{\citenamefont {Born}(1938)}]{Born:1938}%
  \BibitemOpen
  \bibfield  {author} {\bibinfo {author} {\bibfnamefont {M.}~\bibnamefont
  {Born}},\ }\href {\doibase 10.1098/rspa.1938.0060} {\bibfield  {journal}
  {\bibinfo  {journal} {Proceedings of the Royal Society of London. Series A.
  Mathematical and Physical Sciences}\ }\textbf {\bibinfo {volume} {165}},\
  \bibinfo {pages} {291} (\bibinfo {year} {1938})}\BibitemShut {NoStop}%
\bibitem [{\citenamefont {Amelino-Camelia}\ \emph {et~al.}(2011)\citenamefont
  {Amelino-Camelia}, \citenamefont {Freidel}, \citenamefont
  {Kowalski-Glikman},\ and\ \citenamefont {Smolin}}]{AmelinoCamelia:2011bm}%
  \BibitemOpen
  \bibfield  {author} {\bibinfo {author} {\bibfnamefont {G.}~\bibnamefont
  {Amelino-Camelia}}, \bibinfo {author} {\bibfnamefont {L.}~\bibnamefont
  {Freidel}}, \bibinfo {author} {\bibfnamefont {J.}~\bibnamefont
  {Kowalski-Glikman}}, \ and\ \bibinfo {author} {\bibfnamefont
  {L.}~\bibnamefont {Smolin}},\ }\href {\doibase 10.1103/PhysRevD.84.084010}
  {\bibfield  {journal} {\bibinfo  {journal} {Phys. Rev.}\ }\textbf {\bibinfo
  {volume} {D84}},\ \bibinfo {pages} {084010} (\bibinfo {year} {2011})},\
  \Eprint {http://arxiv.org/abs/1101.0931} {arXiv:1101.0931 [hep-th]}
  \BibitemShut {NoStop}%
\bibitem [{\citenamefont {Lobo}\ and\ \citenamefont
  {Palmisano}(2016)}]{Lobo:2016blj}%
  \BibitemOpen
  \bibfield  {author} {\bibinfo {author} {\bibfnamefont {I.~P.}\ \bibnamefont
  {Lobo}}\ and\ \bibinfo {author} {\bibfnamefont {G.}~\bibnamefont
  {Palmisano}},\ }\bibfield  {booktitle} {\emph {\bibinfo {booktitle}
  {{Proceedings, 9th Alexander Friedmann International Seminar on Gravitation
  and Cosmology and 3rd Satellite Symposium on the Casimir Effect: St.
  Petersburg, Russia, June 21-27, 2015}}},\ }\href {\doibase
  10.1142/S2010194516601265} {\bibfield  {journal} {\bibinfo  {journal} {Int.
  J. Mod. Phys. Conf. Ser.}\ }\textbf {\bibinfo {volume} {41}},\ \bibinfo
  {pages} {1660126} (\bibinfo {year} {2016})},\ \Eprint
  {http://arxiv.org/abs/1612.00326} {arXiv:1612.00326 [hep-th]} \BibitemShut
  {NoStop}%
\bibitem [{\citenamefont {Carmona}\ \emph {et~al.}(2019)\citenamefont
  {Carmona}, \citenamefont {Cortés},\ and\ \citenamefont
  {Relancio}}]{Carmona:2019fwf}%
  \BibitemOpen
  \bibfield  {author} {\bibinfo {author} {\bibfnamefont {J.~M.}\ \bibnamefont
  {Carmona}}, \bibinfo {author} {\bibfnamefont {J.~L.}\ \bibnamefont
  {Cortés}}, \ and\ \bibinfo {author} {\bibfnamefont {J.~J.}\ \bibnamefont
  {Relancio}},\ }\href {\doibase 10.1103/PhysRevD.100.104031} {\bibfield
  {journal} {\bibinfo  {journal} {Phys. Rev.}\ }\textbf {\bibinfo {volume}
  {D100}},\ \bibinfo {pages} {104031} (\bibinfo {year} {2019})},\ \Eprint
  {http://arxiv.org/abs/1907.12298} {arXiv:1907.12298 [hep-th]} \BibitemShut
  {NoStop}%
\bibitem [{\citenamefont {Lukierski}\ \emph {et~al.}(1991)\citenamefont
  {Lukierski}, \citenamefont {Ruegg}, \citenamefont {Nowicki},\ and\
  \citenamefont {Tolstoi}}]{Lukierski:1991pn}%
  \BibitemOpen
  \bibfield  {author} {\bibinfo {author} {\bibfnamefont {J.}~\bibnamefont
  {Lukierski}}, \bibinfo {author} {\bibfnamefont {H.}~\bibnamefont {Ruegg}},
  \bibinfo {author} {\bibfnamefont {A.}~\bibnamefont {Nowicki}}, \ and\
  \bibinfo {author} {\bibfnamefont {V.~N.}\ \bibnamefont {Tolstoi}},\ }\href
  {\doibase 10.1016/0370-2693(91)90358-W} {\bibfield  {journal} {\bibinfo
  {journal} {Phys. Lett.}\ }\textbf {\bibinfo {volume} {B264}},\ \bibinfo
  {pages} {331} (\bibinfo {year} {1991})}\BibitemShut {NoStop}%
\bibitem [{\citenamefont {Kowalski-Glikman}\ and\ \citenamefont
  {Nowak}(2003)}]{KowalskiGlikman:2002jr}%
  \BibitemOpen
  \bibfield  {author} {\bibinfo {author} {\bibfnamefont {J.}~\bibnamefont
  {Kowalski-Glikman}}\ and\ \bibinfo {author} {\bibfnamefont {S.}~\bibnamefont
  {Nowak}},\ }\href {\doibase 10.1142/S0218271803003050} {\bibfield  {journal}
  {\bibinfo  {journal} {Int. J. Mod. Phys.}\ }\textbf {\bibinfo {volume}
  {D12}},\ \bibinfo {pages} {299} (\bibinfo {year} {2003})},\ \Eprint
  {http://arxiv.org/abs/hep-th/0204245} {arXiv:hep-th/0204245 [hep-th]}
  \BibitemShut {NoStop}%
\bibitem [{\citenamefont {Kostelecky}(2011)}]{Kostelecky:2011qz}%
  \BibitemOpen
  \bibfield  {author} {\bibinfo {author} {\bibfnamefont {A.}~\bibnamefont
  {Kostelecky}},\ }\href {\doibase 10.1016/j.physletb.2011.05.041} {\bibfield
  {journal} {\bibinfo  {journal} {Phys. Lett.}\ }\textbf {\bibinfo {volume}
  {B701}},\ \bibinfo {pages} {137} (\bibinfo {year} {2011})},\ \Eprint
  {http://arxiv.org/abs/1104.5488} {arXiv:1104.5488 [hep-th]} \BibitemShut
  {NoStop}%
\bibitem [{\citenamefont {Barcelo}\ \emph {et~al.}(2002)\citenamefont
  {Barcelo}, \citenamefont {Liberati},\ and\ \citenamefont
  {Visser}}]{Barcelo:2001cp}%
  \BibitemOpen
  \bibfield  {author} {\bibinfo {author} {\bibfnamefont {C.}~\bibnamefont
  {Barcelo}}, \bibinfo {author} {\bibfnamefont {S.}~\bibnamefont {Liberati}}, \
  and\ \bibinfo {author} {\bibfnamefont {M.}~\bibnamefont {Visser}},\ }\href
  {\doibase 10.1088/0264-9381/19/11/314} {\bibfield  {journal} {\bibinfo
  {journal} {Class. Quant. Grav.}\ }\textbf {\bibinfo {volume} {19}},\ \bibinfo
  {pages} {2961} (\bibinfo {year} {2002})},\ \Eprint
  {http://arxiv.org/abs/gr-qc/0111059} {arXiv:gr-qc/0111059 [gr-qc]}
  \BibitemShut {NoStop}%
\bibitem [{\citenamefont {Weinfurtner}\ \emph {et~al.}(2007)\citenamefont
  {Weinfurtner}, \citenamefont {Liberati},\ and\ \citenamefont
  {Visser}}]{Weinfurtner:2006wt}%
  \BibitemOpen
  \bibfield  {author} {\bibinfo {author} {\bibfnamefont {S.}~\bibnamefont
  {Weinfurtner}}, \bibinfo {author} {\bibfnamefont {S.}~\bibnamefont
  {Liberati}}, \ and\ \bibinfo {author} {\bibfnamefont {M.}~\bibnamefont
  {Visser}},\ }\bibfield  {booktitle} {\emph {\bibinfo {booktitle}
  {{Proceedings, International Workshop on Quantum Simulations via Analogues:
  Dresden, Germany, July 25-28, 2005}}},\ }\href {\doibase
  10.1007/3-540-70859-6_6} {\bibfield  {journal} {\bibinfo  {journal} {Lect.
  Notes Phys.}\ }\textbf {\bibinfo {volume} {718}},\ \bibinfo {pages} {115}
  (\bibinfo {year} {2007})},\ \Eprint {http://arxiv.org/abs/gr-qc/0605121}
  {arXiv:gr-qc/0605121 [gr-qc]} \BibitemShut {NoStop}%
\bibitem [{\citenamefont {{Miron}}(2012)}]{2012arXiv1203.4101M}%
  \BibitemOpen
  \bibfield  {author} {\bibinfo {author} {\bibfnamefont {R.}~\bibnamefont
  {{Miron}}},\ }\href@noop {} {\bibfield  {journal} {\bibinfo  {journal} {arXiv
  e-prints}\ ,\ \bibinfo {eid} {arXiv:1203.4101}} (\bibinfo {year} {2012})},\
  \Eprint {http://arxiv.org/abs/1203.4101} {arXiv:1203.4101 [math.DG]}
  \BibitemShut {NoStop}%
\bibitem [{\citenamefont {Hasse}\ and\ \citenamefont
  {Perlick}(2019)}]{Hasse:2019zqi}%
  \BibitemOpen
  \bibfield  {author} {\bibinfo {author} {\bibfnamefont {W.}~\bibnamefont
  {Hasse}}\ and\ \bibinfo {author} {\bibfnamefont {V.}~\bibnamefont
  {Perlick}},\ }\href@noop {} {\  (\bibinfo {year} {2019})},\ \Eprint
  {http://arxiv.org/abs/1904.08521} {arXiv:1904.08521 [gr-qc]} \BibitemShut
  {NoStop}%
\bibitem [{\citenamefont {Stavrinos}\ and\ \citenamefont
  {Alexiou}(2017)}]{Stavrinos:2016xyg}%
  \BibitemOpen
  \bibfield  {author} {\bibinfo {author} {\bibfnamefont {P.~C.}\ \bibnamefont
  {Stavrinos}}\ and\ \bibinfo {author} {\bibfnamefont {M.}~\bibnamefont
  {Alexiou}},\ }\href {\doibase 10.1142/S0219887818500391} {\bibfield
  {journal} {\bibinfo  {journal} {Int. J. Geom. Meth. Mod. Phys.}\ }\textbf
  {\bibinfo {volume} {15}},\ \bibinfo {pages} {1850039} (\bibinfo {year}
  {2017})},\ \Eprint {http://arxiv.org/abs/1612.04554} {arXiv:1612.04554
  [gr-qc]} \BibitemShut {NoStop}%
\bibitem [{\citenamefont {Girelli}\ \emph {et~al.}(2007)\citenamefont
  {Girelli}, \citenamefont {Liberati},\ and\ \citenamefont
  {Sindoni}}]{Girelli:2006fw}%
  \BibitemOpen
  \bibfield  {author} {\bibinfo {author} {\bibfnamefont {F.}~\bibnamefont
  {Girelli}}, \bibinfo {author} {\bibfnamefont {S.}~\bibnamefont {Liberati}}, \
  and\ \bibinfo {author} {\bibfnamefont {L.}~\bibnamefont {Sindoni}},\ }\href
  {\doibase 10.1103/PhysRevD.75.064015} {\bibfield  {journal} {\bibinfo
  {journal} {Phys. Rev.}\ }\textbf {\bibinfo {volume} {D75}},\ \bibinfo {pages}
  {064015} (\bibinfo {year} {2007})},\ \Eprint
  {http://arxiv.org/abs/gr-qc/0611024} {arXiv:gr-qc/0611024 [gr-qc]}
  \BibitemShut {NoStop}%
\bibitem [{\citenamefont {Amelino-Camelia}\ \emph {et~al.}(2014)\citenamefont
  {Amelino-Camelia}, \citenamefont {Barcaroli}, \citenamefont {Gubitosi},
  \citenamefont {Liberati},\ and\ \citenamefont
  {Loret}}]{Amelino-Camelia:2014rga}%
  \BibitemOpen
  \bibfield  {author} {\bibinfo {author} {\bibfnamefont {G.}~\bibnamefont
  {Amelino-Camelia}}, \bibinfo {author} {\bibfnamefont {L.}~\bibnamefont
  {Barcaroli}}, \bibinfo {author} {\bibfnamefont {G.}~\bibnamefont {Gubitosi}},
  \bibinfo {author} {\bibfnamefont {S.}~\bibnamefont {Liberati}}, \ and\
  \bibinfo {author} {\bibfnamefont {N.}~\bibnamefont {Loret}},\ }\href
  {\doibase 10.1103/PhysRevD.90.125030} {\bibfield  {journal} {\bibinfo
  {journal} {Phys. Rev.}\ }\textbf {\bibinfo {volume} {D90}},\ \bibinfo {pages}
  {125030} (\bibinfo {year} {2014})},\ \Eprint {http://arxiv.org/abs/1407.8143}
  {arXiv:1407.8143 [gr-qc]} \BibitemShut {NoStop}%
\bibitem [{\citenamefont {Letizia}\ and\ \citenamefont
  {Liberati}(2017)}]{Letizia:2016lew}%
  \BibitemOpen
  \bibfield  {author} {\bibinfo {author} {\bibfnamefont {M.}~\bibnamefont
  {Letizia}}\ and\ \bibinfo {author} {\bibfnamefont {S.}~\bibnamefont
  {Liberati}},\ }\href {\doibase 10.1103/PhysRevD.95.046007} {\bibfield
  {journal} {\bibinfo  {journal} {Phys. Rev.}\ }\textbf {\bibinfo {volume}
  {D95}},\ \bibinfo {pages} {046007} (\bibinfo {year} {2017})},\ \Eprint
  {http://arxiv.org/abs/1612.03065} {arXiv:1612.03065 [gr-qc]} \BibitemShut
  {NoStop}%
\bibitem [{\citenamefont {Barcaroli}\ \emph {et~al.}(2015)\citenamefont
  {Barcaroli}, \citenamefont {Brunkhorst}, \citenamefont {Gubitosi},
  \citenamefont {Loret},\ and\ \citenamefont {Pfeifer}}]{Barcaroli:2015xda}%
  \BibitemOpen
  \bibfield  {author} {\bibinfo {author} {\bibfnamefont {L.}~\bibnamefont
  {Barcaroli}}, \bibinfo {author} {\bibfnamefont {L.~K.}\ \bibnamefont
  {Brunkhorst}}, \bibinfo {author} {\bibfnamefont {G.}~\bibnamefont
  {Gubitosi}}, \bibinfo {author} {\bibfnamefont {N.}~\bibnamefont {Loret}}, \
  and\ \bibinfo {author} {\bibfnamefont {C.}~\bibnamefont {Pfeifer}},\ }\href
  {\doibase 10.1103/PhysRevD.92.084053} {\bibfield  {journal} {\bibinfo
  {journal} {Phys. Rev.}\ }\textbf {\bibinfo {volume} {D92}},\ \bibinfo {pages}
  {084053} (\bibinfo {year} {2015})},\ \Eprint
  {http://arxiv.org/abs/1507.00922} {arXiv:1507.00922 [gr-qc]} \BibitemShut
  {NoStop}%
\bibitem [{\citenamefont {Borowiec}\ and\ \citenamefont
  {Pachol}(2010)}]{Borowiec2010}%
  \BibitemOpen
  \bibfield  {author} {\bibinfo {author} {\bibfnamefont {A.}~\bibnamefont
  {Borowiec}}\ and\ \bibinfo {author} {\bibfnamefont {A.}~\bibnamefont
  {Pachol}},\ }\href {\doibase 10.1088/1751-8113/43/4/045203} {\bibfield
  {journal} {\bibinfo  {journal} {J. Phys.}\ }\textbf {\bibinfo {volume}
  {A43}},\ \bibinfo {pages} {045203} (\bibinfo {year} {2010})},\ \Eprint
  {http://arxiv.org/abs/0903.5251} {arXiv:0903.5251 [hep-th]} \BibitemShut
  {NoStop}%
\bibitem [{\citenamefont {Kifune}(1999)}]{Kifune:1999ex}%
  \BibitemOpen
  \bibfield  {author} {\bibinfo {author} {\bibfnamefont {T.}~\bibnamefont
  {Kifune}},\ }\href {\doibase 10.1086/312057} {\bibfield  {journal} {\bibinfo
  {journal} {Astrophys. J.}\ }\textbf {\bibinfo {volume} {518}},\ \bibinfo
  {pages} {L21} (\bibinfo {year} {1999})},\ \Eprint
  {http://arxiv.org/abs/astro-ph/9904164} {arXiv:astro-ph/9904164 [astro-ph]}
  \BibitemShut {NoStop}%
\bibitem [{\citenamefont {Dubovsky}\ and\ \citenamefont
  {Sibiryakov}(2006)}]{Dubovsky:2006vk}%
  \BibitemOpen
  \bibfield  {author} {\bibinfo {author} {\bibfnamefont {S.~L.}\ \bibnamefont
  {Dubovsky}}\ and\ \bibinfo {author} {\bibfnamefont {S.~M.}\ \bibnamefont
  {Sibiryakov}},\ }\href {\doibase 10.1016/j.physletb.2006.05.074} {\bibfield
  {journal} {\bibinfo  {journal} {Phys. Lett.}\ }\textbf {\bibinfo {volume}
  {B638}},\ \bibinfo {pages} {509} (\bibinfo {year} {2006})},\ \Eprint
  {http://arxiv.org/abs/hep-th/0603158} {arXiv:hep-th/0603158 [hep-th]}
  \BibitemShut {NoStop}%
\bibitem [{\citenamefont {Bhattacharya}\ \emph {et~al.}(2012)\citenamefont
  {Bhattacharya}, \citenamefont {Ghrist},\ and\ \citenamefont
  {Kumar}}]{Bhattacharya2012RelationshipBG}%
  \BibitemOpen
  \bibfield  {author} {\bibinfo {author} {\bibfnamefont {S.}~\bibnamefont
  {Bhattacharya}}, \bibinfo {author} {\bibfnamefont {R.}~\bibnamefont
  {Ghrist}}, \ and\ \bibinfo {author} {\bibfnamefont {V.}~\bibnamefont
  {Kumar}}\ }(\bibinfo {year} {2012})\BibitemShut {NoStop}%
\bibitem [{\citenamefont {Petersen}(2006)}]{Petersen2006}%
  \BibitemOpen
  \bibfield  {author} {\bibinfo {author} {\bibfnamefont {P.}~\bibnamefont
  {Petersen}},\ }\href {https://books.google.es/books?id=9cekXdo52hEC} {\emph
  {\bibinfo {title} {Riemannian Geometry}}},\ Graduate Texts in Mathematics\
  (\bibinfo  {publisher} {Springer New York},\ \bibinfo {year}
  {2006})\BibitemShut {NoStop}%
\bibitem [{\citenamefont {Burns}\ \emph {et~al.}(1985)\citenamefont {Burns},
  \citenamefont {Dubrovin}, \citenamefont {Fomenko},\ and\ \citenamefont
  {Novikov}}]{Burns1985}%
  \BibitemOpen
  \bibfield  {author} {\bibinfo {author} {\bibfnamefont {R.}~\bibnamefont
  {Burns}}, \bibinfo {author} {\bibfnamefont {B.}~\bibnamefont {Dubrovin}},
  \bibinfo {author} {\bibfnamefont {A.}~\bibnamefont {Fomenko}}, \ and\
  \bibinfo {author} {\bibfnamefont {S.}~\bibnamefont {Novikov}},\ }\href
  {https://books.google.es/books?id=tlzc7xXYKd8C} {\emph {\bibinfo {title}
  {Modern Geometry— Methods and Applications: Part II: The Geometry and
  Topology of Manifolds}}},\ Graduate Texts in Mathematics\ (\bibinfo
  {publisher} {Springer New York},\ \bibinfo {year} {1985})\BibitemShut
  {NoStop}%
\bibitem [{\citenamefont {Kowalski-Glikman}\ and\ \citenamefont
  {Nowak}(2002)}]{KowalskiGlikman:2002we}%
  \BibitemOpen
  \bibfield  {author} {\bibinfo {author} {\bibfnamefont {J.}~\bibnamefont
  {Kowalski-Glikman}}\ and\ \bibinfo {author} {\bibfnamefont {S.}~\bibnamefont
  {Nowak}},\ }\href {\doibase 10.1016/S0370-2693(02)02063-4} {\bibfield
  {journal} {\bibinfo  {journal} {Phys. Lett.}\ }\textbf {\bibinfo {volume}
  {B539}},\ \bibinfo {pages} {126} (\bibinfo {year} {2002})},\ \Eprint
  {http://arxiv.org/abs/hep-th/0203040} {arXiv:hep-th/0203040 [hep-th]}
  \BibitemShut {NoStop}%
\bibitem [{\citenamefont {Weinberg}(1972)}]{Weinberg:1972kfs}%
  \BibitemOpen
  \bibfield  {author} {\bibinfo {author} {\bibfnamefont {S.}~\bibnamefont
  {Weinberg}},\ }\href
  {http://www-spires.fnal.gov/spires/find/books/www?cl=QC6.W431} {\emph
  {\bibinfo {title} {{Gravitation and Cosmology}}}}\ (\bibinfo  {publisher}
  {John Wiley and Sons},\ \bibinfo {address} {New York},\ \bibinfo {year}
  {1972})\BibitemShut {NoStop}%
\bibitem [{\citenamefont {Poisson}(2009)}]{Poisson:2009pwt}%
  \BibitemOpen
  \bibfield  {author} {\bibinfo {author} {\bibfnamefont {E.}~\bibnamefont
  {Poisson}},\ }\href {\doibase 10.1017/CBO9780511606601} {\emph {\bibinfo
  {title} {{A Relativist's Toolkit: The Mathematics of Black-Hole
  Mechanics}}}}\ (\bibinfo  {publisher} {Cambridge University Press},\ \bibinfo
  {year} {2009})\BibitemShut {NoStop}%
\bibitem [{\citenamefont {Landau}\ and\ \citenamefont
  {Lifschits}(1975)}]{Landau:1982dva}%
  \BibitemOpen
  \bibfield  {author} {\bibinfo {author} {\bibfnamefont {L.~D.}\ \bibnamefont
  {Landau}}\ and\ \bibinfo {author} {\bibfnamefont {E.~M.}\ \bibnamefont
  {Lifschits}},\ }\href@noop {} {\emph {\bibinfo {title} {{The Classical Theory
  of Fields}}}},\ \bibinfo {series} {Course of Theoretical Physics}, Vol.\
  \bibinfo {volume} {Volume 2}\ (\bibinfo  {publisher} {Pergamon Press},\
  \bibinfo {address} {Oxford},\ \bibinfo {year} {1975})\BibitemShut {NoStop}%
\bibitem [{\citenamefont {Barausse}\ \emph {et~al.}(2011)\citenamefont
  {Barausse}, \citenamefont {Jacobson},\ and\ \citenamefont
  {Sotiriou}}]{Barausse:2011pu}%
  \BibitemOpen
  \bibfield  {author} {\bibinfo {author} {\bibfnamefont {E.}~\bibnamefont
  {Barausse}}, \bibinfo {author} {\bibfnamefont {T.}~\bibnamefont {Jacobson}},
  \ and\ \bibinfo {author} {\bibfnamefont {T.~P.}\ \bibnamefont {Sotiriou}},\
  }\href {\doibase 10.1103/PhysRevD.83.124043} {\bibfield  {journal} {\bibinfo
  {journal} {Phys. Rev.}\ }\textbf {\bibinfo {volume} {D83}},\ \bibinfo {pages}
  {124043} (\bibinfo {year} {2011})},\ \Eprint {http://arxiv.org/abs/1104.2889}
  {arXiv:1104.2889 [gr-qc]} \BibitemShut {NoStop}%
\bibitem [{\citenamefont {Blas}\ and\ \citenamefont
  {Sibiryakov}(2011)}]{Blas:2011ni}%
  \BibitemOpen
  \bibfield  {author} {\bibinfo {author} {\bibfnamefont {D.}~\bibnamefont
  {Blas}}\ and\ \bibinfo {author} {\bibfnamefont {S.}~\bibnamefont
  {Sibiryakov}},\ }\href {\doibase 10.1103/PhysRevD.84.124043} {\bibfield
  {journal} {\bibinfo  {journal} {Phys. Rev.}\ }\textbf {\bibinfo {volume}
  {D84}},\ \bibinfo {pages} {124043} (\bibinfo {year} {2011})},\ \Eprint
  {http://arxiv.org/abs/1110.2195} {arXiv:1110.2195 [hep-th]} \BibitemShut
  {NoStop}%
\bibitem [{\citenamefont {Bhattacharyya}\ \emph {et~al.}(2016)\citenamefont
  {Bhattacharyya}, \citenamefont {Colombo},\ and\ \citenamefont
  {Sotiriou}}]{Bhattacharyya:2015gwa}%
  \BibitemOpen
  \bibfield  {author} {\bibinfo {author} {\bibfnamefont {J.}~\bibnamefont
  {Bhattacharyya}}, \bibinfo {author} {\bibfnamefont {M.}~\bibnamefont
  {Colombo}}, \ and\ \bibinfo {author} {\bibfnamefont {T.~P.}\ \bibnamefont
  {Sotiriou}},\ }\href {\doibase 10.1088/0264-9381/33/23/235003} {\bibfield
  {journal} {\bibinfo  {journal} {Class. Quant. Grav.}\ }\textbf {\bibinfo
  {volume} {33}},\ \bibinfo {pages} {235003} (\bibinfo {year} {2016})},\
  \Eprint {http://arxiv.org/abs/1509.01558} {arXiv:1509.01558 [gr-qc]}
  \BibitemShut {NoStop}%
\bibitem [{\citenamefont {Benkel}\ \emph {et~al.}(2018)\citenamefont {Benkel},
  \citenamefont {Bhattacharyya}, \citenamefont {Louko}, \citenamefont
  {Mattingly},\ and\ \citenamefont {Sotiriou}}]{Benkel:2018abt}%
  \BibitemOpen
  \bibfield  {author} {\bibinfo {author} {\bibfnamefont {R.}~\bibnamefont
  {Benkel}}, \bibinfo {author} {\bibfnamefont {J.}~\bibnamefont
  {Bhattacharyya}}, \bibinfo {author} {\bibfnamefont {J.}~\bibnamefont
  {Louko}}, \bibinfo {author} {\bibfnamefont {D.}~\bibnamefont {Mattingly}}, \
  and\ \bibinfo {author} {\bibfnamefont {T.~P.}\ \bibnamefont {Sotiriou}},\
  }\href {\doibase 10.1103/PhysRevD.98.024034} {\bibfield  {journal} {\bibinfo
  {journal} {Phys. Rev.}\ }\textbf {\bibinfo {volume} {D98}},\ \bibinfo {pages}
  {024034} (\bibinfo {year} {2018})},\ \Eprint
  {http://arxiv.org/abs/1803.01624} {arXiv:1803.01624 [gr-qc]} \BibitemShut
  {NoStop}%
\bibitem [{\citenamefont {Cropp}\ \emph {et~al.}(2013)\citenamefont {Cropp},
  \citenamefont {Liberati},\ and\ \citenamefont {Visser}}]{Cropp:2013zxi}%
  \BibitemOpen
  \bibfield  {author} {\bibinfo {author} {\bibfnamefont {B.}~\bibnamefont
  {Cropp}}, \bibinfo {author} {\bibfnamefont {S.}~\bibnamefont {Liberati}}, \
  and\ \bibinfo {author} {\bibfnamefont {M.}~\bibnamefont {Visser}},\ }\href
  {\doibase 10.1088/0264-9381/30/12/125001} {\bibfield  {journal} {\bibinfo
  {journal} {Class. Quant. Grav.}\ }\textbf {\bibinfo {volume} {30}},\ \bibinfo
  {pages} {125001} (\bibinfo {year} {2013})},\ \Eprint
  {http://arxiv.org/abs/1302.2383} {arXiv:1302.2383 [gr-qc]} \BibitemShut
  {NoStop}%
\end{thebibliography}
\end{document}